\documentclass[useAMS,usenatbib]{mn2e}
\usepackage{graphicx}
\usepackage{subfigure}
\usepackage{amssymb}
\usepackage{threeparttable}
\title[BH sub-systems]{Dynamical evolution of black hole sub-systems in idealised star clusters}
\author[P. G. Breen and D. C. Heggie]{ Philip G. Breen$^1$\thanks{E-mail:
p.g.breen@sms.ed.ac.uk} and Douglas C. Heggie$^1$\thanks{E-mail:  d.c.heggie@ed.ac.uk} \\
$^1$ School of Mathematics and Maxwell Institute for Mathematical Sciences, University of Edinburgh, King’s Buildings, Edinburgh EH9 3JZ}

\begin{document}
\date{   \today }
\pagerange{\pageref{firstpage}--\pageref{lastpage}} \pubyear{2012}

\maketitle

\label{firstpage}

\begin{abstract}

In this paper, globular star clusters which contain a sub-system of stellar-mass 
black holes (BH) are investigated. This is done by considering two-component models, 
as these are the simplest approximation of more realistic multi-mass systems, where 
one component represents the BH population and the other represents all the other 
stars. These systems are found to undergo a long phase of evolution where the centre of
 the system is dominated by a dense BH sub-system.  After mass segregation has driven most
 of the BH into a compact sub-system, the evolution of the BH sub-system is found to
 be influenced by the cluster in which it is contained. The BH sub-system evolves in
 such a way as to satisfy the energy demands of the whole cluster, just as
 the core of a one component system must satisfy the energy demands of the whole 
cluster. The BH sub-system is found to exist for a significant amount of time. It 
takes approximately $10t_{\rm rh,i}$, where $t_{\rm rh,i}$ is the initial half-mass relaxation time, 
from the formation of the compact BH sub-system up until the time when $90\%$ of the sub-system total 
mass is lost (which is of order $10^{3}$ times the half-mass relaxation time of the BH sub-system 
at its time of formation). Based on theoretical arguments the rate of mass loss from the 
BH sub-system ($\dot{M}_2$)  is predicted to be $-\beta \zeta M/(\alpha t_{\rm rh} )$,
 where $M$ is the total mass, $t_{\rm rh}$ is the half-mass relaxation time, and $\alpha$, $\beta$, $\zeta$ are three dimensionless parameters
(see Section \ref{sec:BHtwo} for details). An interesting consequence of this is that the rate 
of mass loss from the BH sub-system is approximately independent of the stellar mass ratio 
($m_2/m_1$) and the total mass ratio ($M_2/M_1$)
 (in the range $m_2/m_1 \gtrsim 10$ and $M_2/M_1 \sim 10^{-2}$, where $m_1$, $m_2$ are the masses of individual low-mass and high-mass particles respectively, and $M_1$, $M_2$ are the corresponding total masses). The theory is 
found to be in reasonable agreement with most of the results of a series of N-body simulations, and with
 all of the models if the value of $\zeta$ is suitable adjusted. Predictions based on theoretical arguments 
are also made about the structure of BH sub-systems. Other aspects of the evolution are also considered such as 
the conditions for the onset of gravothermal oscillation.

\end{abstract}


\section{Introduction}

Hundreds of stellar mass black holes (BH) can form within 
a massive globular cluster (see \cite{Kulkarnietal1993}, \cite{Sigurdsson1993} 
and \cite{Port2000}). Some of the BH might escape at the 
time of their formation due to large natal kicks. However 
the subject of natal kicks for BH is still under debate 
\citep{Repettoetal2012} and it is possible that the largest BH may 
form without any supernova explosion \citep{Fryer1999}. 
Uncertainty in the natal kicks leads to uncertainty in 
the initial size of the BH population. As the BH are more massive than the other stars in the system, any retained BH will undergo mass segregation and almost all are likely to become concentrated in the centre of the system, eventually forming a compact sub-system.

The mass of the BH sub-system decreases over time 
because BH binaries form in the dense core of the BH sub-system, 
causing the ejection of single BH and ultimately the 
binaries themselves through super-elastic encounters (see \cite{Kulkarnietal1993}, \cite{Sigurdsson1993}, \cite{Port2000}, \cite{Banerjee2010}, \cite{Downingetal2010}, \cite{aarsethnbody}). Early work by \cite{Kulkarnietal1993} and \cite{Sigurdsson1993} seemed to indicate that the BH population will 
become depleted over a relatively short timescale. This 
conclusion was reached in part by treating the BH sub-system 
as if it were an independent system once most of the BH had 
segregated to the centre of the cluster.

\cite{Merritt2004} and \cite{Mackey2007} found that 
heating by a retained population of BH causes large-scale
 core expansion in massive star clusters. They suggest this
 may partly explain the core radius-age trend observed for such objects in the Magellanic Clouds. The BH binaries that are formed in the core of the BH sub-system are an interesting class of objects in their own right, especially as the merger of two BH may be detectable as a source of 
gravitational waves (\cite{Port2000} and 
\cite{Banerjee2010}). It has even been suggested that star 
clusters consisting almost entirely of BH, known as dark star 
clusters, could exist \citep{Banerjee2011}. Dark star clusters 
could be created if the stars in the outer parts of a 
larger system were stripped away by a strong tidal field, leaving behind the BH sub-system. If one were to observe the few remaining stars in these systems they would appear to be super virial, as the velocity dispersion of the remaining stars would be enhanced by the unseen BH.

\cite{breenheggie1} investigated the evolution of two-component models and found that, within the parameter space they considered, the stability of the two-component system against gravothermal oscillations was dominated by the heavy component. They only considered systems with a total mass ratio of order $M_2/M_1\gtrsim 10^{-1}$, where $M_2$ ($M_1$) is the total mass of the heavy (light) component. However the mass ratio of a system containing a BH sub-system would only be expected to be $M_{2}/{M_1}\sim10^{-2}$ \citep{Port2000}, where $M_{2}$ is the total mass of the BH sub-system, which is smaller by an order of magnitude than any of the systems
 studied by \cite{breenheggie1}. As H\'{e}non's Principle \citep{Henon1975} 
states that the energy generating rate of the 
core is regulated by the bulk of the system, it 
seems unlikely that the approach of \cite{breenheggie1} is appropriate 
in this case due to the small value of $M_2/M_1$.

This paper is structured as follows. In Section \ref{sec:BHtwo}, some theoretical results are derived and discussed. This is followed by Section \ref{sec:rhsec}, where the theoretical results regarding the size of the BH sub-system are tested using both gas models and direct N-body runs. Section \ref{sec:Nbody} contains empirical results regarding the mass loss rates from BH sub-systems and a comparison between the empirical results and the theory of Section \ref{sec:BHtwo}. The qualitative behaviour of these systems is also discussed in this section. Section \ref{sec:GTO} is concerned with gravothermal oscillations in systems containing a BH sub-system. Finally Section \ref{sec:conanddis} consists of the conclusions and a discussion. 


\section{Theoretical Understanding}\label{sec:BHtwo}

\subsection{BH sub-system: half-mass radius}\label{sec:theory}

Here we will consider aspects of the dynamics of a system 
containing a BH sub-system. We will assume that the system is
 Spitzer unstable \citep{Spitzer} and that the total mass of 
the BH sub-system ($M_{2}$) is very small compared to the 
total mass of the system $(M)$. (Since the Spitzer stability criterion is $ (M_2/M_1)(m_2/m_1)^{\frac{3}{2}}< 0.16$, these assumptions are consistent provided the stellar mass ratio $m_2/m_1$ is large enough). We will also assume that the 
initial state of the system has a constant mass density ratio 
between the two components throughout and that the 
velocity dispersions of both components are equal at all locations.
 If this is the case then the system would first experience a 
mass segregation-dominated phase of evolution which lasts 
of order $({m_1}/{m_2})t_{\rm cc}$ \citep[and references therein]{Fregeau}, 
where $m_2$ ($m_1$) is the stellar mass of the BH (other stars), and $t_{\rm cc}$ is the core collapse time in a single component system, although technically for the outermost BH mass segregation can last much longer than $({m_1}/{m_2})t_{\rm cc}$ (see Appendix \ref{sec:Heatingouterlagra} for details).

If we consider the $50\%$ Lagrangian shell of the heavy component, 
initially it will be approximately the same size as the  $50\%$ 
Lagrangian shell of the entire system. As the BH lose energy to 
the other stars in the system the $50\%$ Lagrangian shell of the 
BH  component contracts. The shell will continue to contract until 
the energy loss to the light component is balanced by the energy the 
shell receives from the inner parts of the BH  component. As we have 
assumed that the system is Spitzer unstable, it follows that a 
temperature difference must remain between the two components and 
thus there is still a transfer of energy between the two components. 
As the total mass of BH is small, the contraction of the $50\%$ Lagrangian 
shell of the heavy component continues until the system is concentrated 
in a small region in the centre of the system. This is what we call the BH sub-system. 
The BH sub-system is very compact and therefore rapidly undergoes core 
collapse. The subsequent generation of 
energy by the formation of BH binaries, and interactions of BH binaries 
with single BH, support the $50\%$ Lagrangian shell
 of the heavy component.

In the present paper, we will assume that the main 
pathway for the transport of thermal energy throughout 
the system is as follows: energy is generated in the core of
 the BH sub-system (we are assuming that the BH core 
radius is much smaller than $r_{\rm h,2}$, the half mass 
radius of the BH sub-system), then the energy is conducted 
throughout the BH sub-system via two-body relaxation just as in the conventional picture of post-collapse evolution; but in the standard one-component setting this flux causes expansion and, ultimately, dispersal of the system. In a two-component system, however, the coupling to the lighter component changes the picture dramatically. We will assume that at a radius comparable 
with $r_{\rm h,2}$ most of the energy flux is transferred into 
the light component, where it then spreads throughout the bulk 
of the light system. These assumptions will hold if most of the heating, either direct (heating by reaction products which remain in the cluster) or indirect, initially occurs within the BH sub-system rather than within the regions dominated by the light component (see Appendix \ref{ABH} for a discussion on this issue). A similar assumption is made in one-component gas models \citep{Goodman1987, Heggieramamani}  which have been shown to be in good agreement with direct N-body
 models \citep{betsug1985}.

From H\'{e}non's Principle \citep{Henon1975} we argue that the
 rate of energy generation is regulated by the energy demands
 of the bulk of the system. For the systems we are considering
 here the bulk of the system is in the light component as 
$M_1\approx 0.99M$ for $M_2/M_1=10^{-2}$. The energy demands of a 
system are normally thought of as the energy flux at the half-mass 
radius, which is of order $|E|/t_{\rm rh}$, where $E$ is the total energy of
 the system and $t_{\rm rh}$ is the half mass relaxation time. Under our assumptions the energy flux must be supplied by the BH sub-system, which ultimately must be generated by binaries in the core of the BH sub-system. However, we can ignore the details of how the energy is actually generated and consider the BH sub-system itself as the energy source for the cluster as a whole. In this picture, the energy exchange ($\dot{E}_{\rm ex}$) between the BH sub-system and the light component must balance the flux at the half mass radius of the light 
component i.e. 
$$ \frac{|E|}{t_{\rm rh}} \sim |\dot{E}_{\rm ex}|.$$

The BH sub-system is concentrated in the centre of the cluster, and therefore the half-mass relaxation time in the BH sub-system ($t_{\rm rh,2}$) is quite short. It follows that the flux at the half-mass radius of the BH sub-system is quite high. The flux at the half-mass radius of the BH sub-system ($r_{\rm h,2}$) is of order $|E_{2}|/t_{\rm rh,2}$, where $E_2$ is the energy of the BH sub-system. Most of the energy that passes $r_{\rm h,2}$ must be transferred to the light component or else the BH sub-system would rapidly expand until the flux around $r_{\rm h,2}$ is comparable to  the rate of energy exchange. Therefore the energy exchange rate between the two components must be approximately equal to the flux at the half mass radius of the heavy sub-system i.e.
 
$$  \frac{|E_{2}|}{t_{\rm rh,2}} \approx |\dot{E}_{\rm ex}|.$$

All this leads to the conclusion that the flux at $r_{\rm h}$ must be balanced by the flux at $r_{\rm h,2}$ i.e.
$$ \frac{|E|}{t_{\rm rh}} \sim  \frac{|E_{2}|}{t_{\rm rh,2}},$$
which can be rearranged as 
\begin{center}
\begin{equation}\label{eq:enerel}
\; \; \; \; \; \; \;\; \; \; \; \; \; \;\; \; \; \; \; \; \; \; \; \; \; \; \; \;\; \; \; \; \; \; \;\; \; \; \frac{|E|}{|E_{2}|} \sim \frac{t_{\rm rh}}{t_{\rm rh,2}}. 
\end{equation}
\end{center}
Using the definition of $t_{\rm rh}$ as given in \cite{Spitzer} (and an equivalent definition for $t_{\rm rh,2}$), the right hand side of equation \ref{eq:enerel} becomes

\begin{center}
\begin{equation} \label{eq:bal}
\; \; \; \; \; \; \;\; \; \; \; \frac{t_{\rm rh}}{t_{\rm rh,2}} \approx \frac{N^\frac{1}{2} r_{\rm h}^\frac{3}{2}m_{2}^\frac{1}{2}\ln{\Lambda_2}}{N_{2}^\frac{1}{2} r_{\rm h,2}^\frac{3}{2} m^\frac{1}{2} \ln{\Lambda}}=\frac{ M^\frac{1}{2}  r_{\rm h}^\frac{3}{2} m_{2} \ln{\Lambda_2} }{ M_{2}^\frac{1}{2}  r_{\rm h,2}^\frac{3}{2} m\ln{\Lambda}}, 
\end{equation}
\end{center}
where $m$ is the mean mass, and $\ln{\Lambda}$, $\ln{\Lambda_2}$ are the coulomb logarithms of the entirely system and the BH sub-system respectively. 

Using equation \ref{eq:enerel}, the left hand side of equation \ref{eq:bal} becomes 
\begin{center}
\begin{equation} \label{eq:bal2}
 \; \; \; \; \; \; \;\; \; \; \; \; \; \;\; \; \; \; \; \; \; \;\; \; \; \; \; \; \;\; \; \;\frac{|E|}{|E_{2}|} \approx \frac{M\sigma^2}{M_{2}\sigma_{2}^2} \approx  \frac{M^2 r_{\rm h,2}}{M_{2}^2r_{\rm h}}. 
\end{equation}
\end{center}
where we have estimated the squared one dimensional velocity dispersions, $\sigma_{2}^2$ and $\sigma^2$, by assuming that both the system and the sub-system are in virial equilibrium, so that $\sigma^2\simeq 0.2GM/r_{\rm h}$ and $\sigma_2^2\simeq 0.2GM_2/r_{\rm h,2}$. Putting the above equations together we have
$$ \frac{M^2 r_{\rm h,2}}{M_{2}^2r_{\rm h}}   \sim  \frac{M^\frac{1}{2} m_{2} r_{\rm h}^\frac{3}{2}\ln{\Lambda_2}}{M_{2}^\frac{1}{2} m r_{\rm h,2}^\frac{3}{2}\ln{\Lambda}} $$ and then by rearranging this we have
\begin{center}
\begin{equation} \label{eq:fin}
 \; \; \; \; \; \; \;\; \; \; \; \; \; \;\; \; \; \; \; \; \; \;\; \; \; \; \; \; \;\; \; \;\frac{r_{\rm h,2}^\frac{5}{2}}{r_{\rm h}^\frac{5}{2}}  \sim \frac{M_{2}^\frac{3}{2}}{M^\frac{3}{2}} \frac{m_{2}}{m} \frac{\ln{\Lambda_2}}{\ln{\Lambda}}.
\end{equation}
\end{center}

This result implies that for a fixed total mass ratio $M_{2}/M_1$ ($\approx M_2/M$) and ignoring the variation of the coulomb logarithms, the ratio of $r_{\rm h,2}/r_{\rm h}$ grows with increasing stellar mass ratio $m_{2}/m$.  
\subsection{BH sub-system: core radius}\label{sec:bal}

In Section \ref{sec:theory} $r_{\rm h,2}/r_{\rm h}$ was estimated by assuming that the energy flow in the BH sub-system balances the energy flow in the bulk of the system (i.e. the other stars). In order for equation \ref{eq:fin} to hold it is assumed that the BH core radius $r_{\rm c,2}$ must be significantly smaller than $r_{\rm h,2}$ and that the BH sub-system is actually capable of producing and supplying the energy required by the system. Usually it is assumed that the core (in this case the core of the BH sub-system) adjusts to provide the energy required. In this section we estimate the size of the core, and use the estimate to place a condition on the validity of our assumptions. 

As most of the mass within $r_{\rm h,2}$ is BH, we may treat the 
BH sub-system as a one-component system and we can make 
use of the standard treatments of one-component systems. In 
balanced evolution (i.e. a situation in which energy is produced 
at the rate at which it flows over the half-mass radius), 
the rate of energy production is given by $\dot{E} = (|E_2|/t_{\rm rh,2}) \zeta_2$, where 
$\zeta_2$ is a constant (for a one-component model $\zeta_2\approx 0.0926$, 
see \cite{Henon}). We will follow the derivation in  \citet[][page 265]{HeggieHut2003} 
of the dependence of $r_{\rm c}/r_{\rm h}$ on $N$ for 
a one component model, although here it will be necessary
 to keep track of the numerical constants and account for 
the fact that the properties correspond to those of the BH sub-system. In order to derive a condition on $r_{\rm c,2}/r_{\rm h,2}$ it is necessary to express $\dot{E}$, $E_2$ and $t_{\rm rh,2}$ in terms of the other properties of the BH sub-system. $\dot{E} \approx M_{\rm c,2} \epsilon$  where $M_{\rm c,2}$ is the BH core mass and $\epsilon$ is the energy generating rate per unit mass \citep{HeggieHut2003}. $M_{\rm c,2}$ and $\epsilon$ can be expressed in terms of $\rho_{\rm c,2}$ (the central mass density 
of BH),
 $r_{\rm c,2}$ and $\sigma_{\rm c,2}$ (the central one-dimensional velocity dispersion of the BH), which results in
$$\dot{E} = 85\frac{G^5 m_2^3 \rho_{\rm c,2}^3 r_{\rm c,2}^3}{\sigma_{\rm c,2}^{7}}.$$ 
As in the previous section we shall use $$|E_2| \approx 0.2 \frac{GM_2^2}{r_{\rm h,2}}.$$ 
It will be convenient to use a different but equivalent definition of $t_{\rm rh,2}$ \citep{Spitzer} rather than the one used in the previous section, i.e.
$$t_{\rm rh,2} =\frac{0.195\sigma_2^3}{Gm_2\rho_{\rm h,2}\ln{\Lambda_2}},$$
 where $\sigma_2$ is the one-dimensional velocity dispersion of the BH inside $r_{\rm h,2}$ and $\rho_{\rm h,2}$ is the mean mass density of the BH inside $r_{\rm h,2}$.

Using $\dot{E} = (|E_2|/t_{\rm rh,2} ) \zeta_2$ all of the above equations can be combined into 
$$ 83 G^2 m_2^2 \rho_{\rm c,2}^3\sigma_{2}^3 r_{\rm c,2}^3 r_{\rm h,2} \approx M_2^2 \sigma_{\rm c,2}^7\rho_{\rm h,2} \zeta_2\ln{\Lambda_2}.$$
This can be simplified by using $4\pi G \rho_{\rm c,2}r_{\rm c,2}^2 = 9\sigma_{\rm c,2}^2$. Also the BH sub-system is expected to be nearly isothermal inside $r_{\rm h,2}$;  therefore $\sigma_{2} \approx \sigma_{\rm c,2}$ and furthermore it follows that $\rho_{2}\propto r^{-2}$ inside $r_{\rm h,2}$. Taking all of this into account and by rearranging the above we have 

\begin{equation}\label{eq:balone}
  \; \; \; \; \; \; \;\; \; \; \; \; \; \;\; \; \; \; \;\; \; \; \;\frac{r_{\rm c,2}}{r_{\rm h,2}} \approx N_2^{-\frac{2}{3}} \big(\frac{104}{\zeta_2\ln{ \Lambda_2}}\big)^{\frac{1}{3}}.   
\end{equation}

\subsection{Limitations of the theory}\label{sec:limitations}
One of our assumptions was that $r_{\rm c,2} \ll r_{\rm h,2}$ and now we can derive an  approximate condition for the validity of the theory. As $N_2$ is small we shall take $\ln{\Lambda_2} \approx 1$. As the entire system is in balanced evolution it is also true that $(|E_2|/t_{\rm rh,2}) \zeta_2 = (|E|/t_{\rm rh}) \zeta$, where $\zeta$ is a dimensionless parameter defined implicitly by the equation $\dot{E}=(|E|/t_{\rm rh}) \zeta$; we expect $|E_2|/t_{\rm rh,2} \sim |E|/t_{\rm rh}$ (equation \ref{eq:enerel}) and so for the purposes of our estimate we can assume that $\zeta_2 \approx \zeta$. Therefore  $r_{\rm c,2} \lesssim r_{\rm h,2}$ provided $N_2 \gtrsim 40$. This value is only a rough guide, and what is important to take from this result is that for sufficiently small $N_2$ the theory in Section \ref{sec:theory} breaks down. 

As $M_2$ decreases the BH sub-system will ultimately reach a point where it can no longer solely power the expansion of the system by the mechanism we have considered (i.e. formation, hardening and ejection of BH binaries by interaction amongst the BH sub-system). After this point 
it may be possible for BH binaries to generate the required
 energy through strong interaction with the light stars. However
 this will probably require a much higher central mass 
density of the light component than at the time of formation of the BH sub-system, as at this central mass 
density interactions between the light stars and the BH binaries are expected to be much less efficient at generating energy than interactions between single BH and BH binaries. This implies a potentially significant adjustment phase towards the end of the life of the BH sub-system, as is illustrated by an N-body model in Section \ref{sec:Nbody} (see Fig. \ref{fig:BEHA}).

In Section \ref{sec:theory} we made the assumption that the BH sub-system was Spitzer stable. However, as pointed out to us by Sambaran Banerjee (private communication), it is also possible that as $M_2$ decreases a point may be reached were the sub-system becomes Spitzer stable. If so, the two components could reach equipartition at the centre, i.e. $m_2\sigma_2^2=m_1\sigma_1^2$, and in that case our assumption that heat flows from the heavy component to the light component is false. The temperature ratio of the two component may be estimated by
$$\frac{m_2\sigma_2^2}{m_1\sigma_1^2} \sim \frac{m_2}{m_1}\frac{M_2}{M_1} \frac{r_{\rm h}}{r_{\rm h,2}} \sim \big(\frac{m_2}{m_1} \big)^{\frac{3}{5}} \big(\frac{M_2}{M_1} \big)^{\frac{2}{5}}  \big(\frac{\ln{\Lambda_2}}{\ln{\Lambda}}\big)^{-\frac{2}{5}},$$
were we have made use of equation \ref{eq:fin} and the assumptions made in Section \ref{sec:theory}. Initially $(m_2\sigma_2^2)/(m_1\sigma_1^2) >  1$ but for fixed $m_2/m_1$ it decreases towards equipartition  ($(m_2\sigma_2^2)/(m_1\sigma_1^2) = 1$) as BH escape and $M_2$ decreases. By setting $(m_2\sigma_2^2)/(m_1\sigma_1^2) = 1$,  ignoring the variation of coulomb logarithms and taking each side to the power $2/5$ we arrive at 
$$ \frac{M_2}{M_1} \sim C \big(\frac{m_2}{m_1} \big)^{-\frac{3}{2}}  $$ 
where $C$ is a constant. This is exactly the same form as the Spitzer stability criterion \citep{Spitzer}, the only difference being that we have not specified the constant $C$. Again it follows that the theory of Sections \ref{sec:theory} and \ref{sec:bal} will fail when $M_2$ becomes too small, and that the limiting value of $M_2$ is smaller for larger $m_2/m_1$.

We will now briefly consider the case of a Spitzer stable BH sub-system. As the BH move more slowly than the other stars they still concentrate in the centre of the system. If the heavy component still dominates within $r_{\rm h,2}$ then the BH sub-system is self-gravitating and
$$\sigma^2_2 \sim \frac{GM_2}{r_{\rm h,2}}$$
still holds. From equipartition of kinetic energy we have
$$ \frac{m_1}{m_2}\sigma^2_1  \sim  \frac{m_1}{m_2} \frac{GM}{r_{\rm h}}  \sim \frac{GM_2}{r_{\rm h,2}}.$$
This result can be rearranged as
\begin{equation}\label{eq:balonestable}
 \hspace{30mm} \frac{r_{\rm h,2}}{r_{\rm h}} \sim \frac{m_2}{m_1}\frac{M_2}{M}.
\end{equation}
In equation \ref{eq:balonestable} there is a different dependence of $r_{\rm h,2}/r_{\rm h}$ on $m_2/m_1$ and $M_2/M$ than in equation \ref{eq:balone}.

\subsection{Evaporation rate}\label{sec:evap}

We will now consider the evaporation rate as a result of two-body 
encounters for the BH sub-system. It is important to note that evaporation is only one of the mechanisms by which BH are removed from the system. Another important mechanism as already discussed is ejection via encounters involving BH binaries and single BH. In this section we will ignore this effect although it will be considered in detail in the next section. 

The one dimensional velocity dispersion of the BH sub-system has the following dependence on $M_{2}$ and $r_{\rm h,2}$ (assuming it is nearly self gravitating): $$\sigma_{2}^2 \sim \frac{GM_{2}}{r_{\rm h,2}}.$$ From the previous section we know that $\displaystyle{\big(\frac{r_{\rm h,2}}{r_{\rm h}}\big)^\frac{5}{2} \sim \frac{M_{2}^\frac{3}{2}}{M^\frac{3}{2}} \frac{m_{2}}{m} \frac{\ln{\Lambda_2}}{\ln{\Lambda}}}$. Therefore, if we consider the post collapse evolution of a series of models with different values of $m_{2}/m$, at the same values of $r_{\rm h}$, $M_2$ and $M$, as $m_{2}/m$ increases so does $r_{\rm h,2}$ and therefore $\sigma_{2}^2$ decreases. Here we have ignored the variation of the coulomb logarithms; for a system with $N=10^6$ and $M_2/M_1=10^{-2}$, the variation of $\ln{\Lambda_2}/\ln{\Lambda}$ is a factor of $2.2$ between $m_2/m_1=10$ and $50$ if $\Lambda_2 = 0.02N_2 $ and $\Lambda = 0.02N$. 
The source of the value $0.02$ is \cite{GierszHeggie2}.

The mean-square escape velocity is related to the mean-square velocity of the system (see \cite{Spitzer} and \cite{GaDy}) by $ v_{\rm e}^2 = 4 v^2.$ The mean-square velocity of the system is dominated by the light component and remains approximately fixed with varying $m_{2}/m$. Therefore, as $m_{2}/m$ increases the mean-square velocity of the BH sub-system decreases relative to the  mean-square escape velocity. This implies that systems with higher $m_{2}/m$ lose a lower fraction of their stars by evaporation per $t_{\rm rh,2}$; in fact we will show in the next paragraph that escape via evaporation is negligible.

A rough estimate of the fraction of stars lost by evaporation each $t_{\rm rh}$ can be calculated from the Maxwellian velocity distribution \citep{Spitzer,GaDy}. This is done by assuming that the fraction of stars with $v$ greater than $v_{\rm e}$ in a Maxwellian velocity distribution is removed each $t_{\rm rh}$, i.e.
$$\frac{dN}{dt}=-\frac{N}{t_{\rm rh}}\gamma,$$ 
where $\gamma$ denotes the fraction of stars removed. Its value for a one component model is $\gamma = 7.38 \times 10^{-3}$. In order to estimate the value of $\gamma$ for a two component model we need to know the relationship between $v_2^2$ and $v_1^2$, where $v_2^2$ ($v_1^2$) is the mean-square velocity of the heavy (light) component. If the system was Spitzer stable then  $m_2 v_2^2 = m_1 v_1^2$; however the systems we are considering are not Spitzer stable because of the large stellar mass ratios, and therefore it is expected that $m_2v_2^2 > m_1 v_1^2$. Over a period of time where $M_2$ and $r_{\rm h}$ remain roughly constant then $ v_2^2$ and $v_1^2$ will be approximately constant. Therefore over the same time period we have $m_2 v_2^2 \approx k m_1 v_1^2$, where $k$ is a constant. Using a two-component gas model (see \cite{HeffieAarseth1992} and \cite{breenheggie1}), for the range of parameters in Section \ref{sec:rhsec}, $k$ was found to be $\le 2$. Assuming a stellar mass ratio of $10$ and letting $k=2$ 
to insure 
the 
highest possible value of $\gamma$ leads to $\gamma = 5.87 \times 10^{-13}$. This exceedingly small value of $\gamma$ is a result of the fact that the Maxwellian velocity distribution drops exponentially with increasing velocity, so that even a slight increase in escape velocity leads to a much smaller value of $\gamma$.

Based on this approximate theory we can conclude that mass loss from evaporation due to two body encounters is not significant for the case of BH sub-systems. It is worth noting (based on the arguments in this section) that constraints based on evaporation timescales \citep[for example see][]{Maoz} which only take into consideration the potential of the BH sub-system are not generally valid if the sub-system is embedded in a much more massive system. BH which escape the sub-system in two-body encounters generally cannot escape from the deep potential well of the surrounding system. Instead, they return to the sub-system on the mass segregation/dynamical friction timescale.

\subsection{Ejection rate}\label{sec:ej}
Dynamical evolution of BH binaries and ejection of BH is an energy source which is assumed in the present paper to comply with H\'{e}non's Principle. As has been stated in Section \ref{sec:theory},  H\'{e}non's Principle states that $\dot{E}$ is regulated by the energy demands of the bulk of the system, i.e. 
\begin{center}
\begin{equation} \label{eq:ETRH_ZETA}
 \; \; \; \; \; \; \;\; \; \; \; \; \; \;\; \; \; \; \; \; \; \;\; \; \; \; \; \; \;\; \; \;\dot{E}\simeq\frac{|E|}{t_{\rm rh}}\zeta,
\end{equation}
\end{center}
where $\zeta$ is a constant. For systems with $M_2 \ll M$, $(|E|/t_{\rm rh})\zeta$ is determined mainly by the properties of the light component and is approximately independent of $M_2$ and $m_2$. Therefore the energy generation rate is also approximately independent of $M_2$ and $m_2$.

The encounters which generate energy (either by formation of binaries or their subsequent harding) happen where the density is highest, in the core of the BH sub-system. As the BH are concentrated in the centre of the system, through mass segregation, we may assume that encounters which generate energy predominantly occur between BH binaries and single BH. The BH sub-systems considered in this paper consist of BH with identical stellar mass, therefore the mechanism by which energy is generated in the BH sub-systems is similar to that for a one-component system. The two key differences for the BH sub-system are that the escape potential is elevated and the size of the system is regulated by the much more massive system of light stars (see equation \ref{eq:fin}). 

In a one-component system each hard binary formed in the core on average contributes a fixed amount of energy $\propto m\phi_{\rm c}$ (where $\phi_{\rm c}$ is the central potential) before being ejected from the system \citep[see for example][]{HeggieHut2003}. Typical estimates of the average energy each hard binary contributes in a one component system are $\approx 7.5m\phi_{\rm c}$ \citep{Goodman1984}  and $\approx 8.27m\phi_{\rm c}$ \citep{HeggieHut2003}\footnote{Note there is an error in \cite{HeggieHut2003} p. 225: the constant is stated incorrectly but the correct value can be obtained by evaluating the formula given on the same page.}. Also on average each hard binary causes the ejection of a fixed number of stars. \cite{Goodman1984} estimated this to be approximately $6$ stars (including the binary itself) and \cite{HeggieHut2003} estimated this to be approximately $3$ stars (excluding the binary itself). The situation is similar for a BH sub-system and we can assume that the mass ejected and the average contribution per hard BH binary is the same as for the 
one-component case. Furthermore as mass loss due to evaporation is negligible for a BH sub-system (see 
Section \ref{sec:evap}), the loss of mass from the sub-system is always associated with energy generation. Therefore we can express the rate of energy generation in the core in terms of mass loss, 
\begin{equation}\label{eq:Egen_beta}
 \; \; \; \; \; \; \;\; \; \; \; \; \; \;\; \; \; \; \; \; \; \;\;\;\;\;\; \;\;\;\;\;\; \;\;\;\;\;\; \dot{E}\approx \beta\dot{M_2}\phi_{\rm c},
\end{equation}
where $\beta$ is a constant; $\beta \approx 2.2$ in the one component case, where we have used the values of energy generated and mass lost given in  \cite{HeggieHut2003}, adjusted to account for the energy generated and mass lost in the escape of the binary itself,  $\approx 10.6 m_2 \phi_{\rm c}$ and $\approx 4.7m_2$ receptively. Since $\dot{E}$ is regulated by the light component (equation \ref{eq:ETRH_ZETA}), we can use equation \ref{eq:Egen_beta} to estimate the rate of mass loss. 
Note that the estimates in this paragraph are entirely theoretical, without detailed numerical support especially for the value of $\beta$. Note also that the estimate ignores the heating effect of encounters which do not lead to ejection once the binary has reached a sufficient binding energy for ejection to be likely.

We will now show that $\phi_{\rm c}$ is approximately independent of the properties of the BH sub-system. This will be done by showing that the main contribution to the central potential is from the light component. We can estimate the contribution of the lights to $\phi_{\rm c}$ to be  $ \phi_1 \approx -GM/r_{\rm h}$ and the contribution of the BH to be $\phi_2 \approx -GM_2/r_{\rm h,2}$. In the regime of interest  $M_2/M=10^{-2}$ and $r_{\rm h,2}/r_{\rm h}=10^{-1}$ (for typical values of $r_{\rm h,2}/r_{\rm h}$: see Table \ref{table:rBHrh}). Therefore,


$$\frac{\phi_2}{\phi_1}\approx \frac{M_2}{M} \frac{r_{\rm h}}{r_{\rm h,2}} \approx 10^{-1},$$ and we can approximate $\phi_{\rm c}$ by $\phi_1$ (see also Section \ref{sec:phi}).

We can now use $\dot{E}\approx \beta\dot{M_2}\phi_{\rm c}$ to make an estimate of the mass loss rate: from equations \ref{eq:ETRH_ZETA} and \ref{eq:Egen_beta} we have

$$ \beta \dot{M_2} \phi_{\rm c} \simeq \frac{|E|}{t_{\rm rh}}\zeta,$$
and so
$$ \dot{M_2}  \simeq M\frac{|E|}{M\phi_{\rm c}}\frac{1}{t_{\rm rh}}  \frac{\zeta}{\beta}.$$
The term $E/(M\phi_{\rm c})$ is dimensionless and approximately independent of the properties of the BH sub-system; we will use $\alpha$ to represent its value. For a Plummer model $\alpha \approx 0.15$, however during core collapse $|\phi_{\rm c}|$ increases while $E$ and $M$ remain approximately constant, resulting in smaller values of $\alpha$. The two-component gas models used in Section \ref{sec:rhsec} indicate a value of $\alpha \approx 0.13$. We now have

\begin{center}
\begin{equation}\label{eq:M2dot_general}
\hspace{2cm} \dot{M_2}  \simeq -\frac{M}{t_{\rm rh}}  \frac{\alpha\zeta}{\beta} 
 \end{equation}
\end{center}
Scaling to the values of $\alpha$, $\zeta$ and $\beta$, we have 
\begin{center}
\begin{equation}\label{eq:M2dot}
\hspace{2cm} \dot{M_2}  \simeq -  0.0061  \frac{M}{t_{\rm rh}}  \frac{\alpha}{0.15}  \frac{\zeta}{0.09}  \frac{2.2}{\beta}.
 \end{equation}
\end{center}

By this estimate the sub-system should last  $\sim 1.6 t_{\rm rh}$ to $3.3t_{\rm rh}$, for $M_2/M_1=0.01$ to $0.02$  and canonical values of $\alpha$, $\beta$ and $\zeta$. The important point to take from this result is that the rate of mass loss from the BH sub-system depends on the half-mass relaxation time of the whole system and not on any property of the BH sub-system.

While a system is in balanced evolution, the only parameter that varies significantly (over a timescale where $\dot{M}$ is negligible) in the right hand side of the above equation is $t_{\rm rh}$, due to the increase in $r_{\rm h}$ (Here we assume that the system is isolated; the case of a tidally limited system is considered in the following section.). Therefore, for a particular system equation \ref{eq:M2dot} can be expressed in the form $ \dot{M_2}  \simeq -Cr_{\rm h}^{-\frac{3}{2}}$, where $C$ is a constant ($\displaystyle{C=\frac{\alpha \zeta}{\beta}\frac{Mr_{\rm h,i}^{\frac{3}{2}}}{t_{\rm rh,i}}}$ where $\alpha \zeta/\beta \approx 6.1\times10^{-3}$ for canonical values of $\alpha$, $\beta$, and $\zeta$ and $i$ denotes values at the start of the balanced evolution). $r_{\rm h}$ itself is a function of time which can be derived from the relation $\dot{r}_{\rm h}/r_{\rm h} = \zeta/t_{\rm rh}$, which follows in turn from equation \ref{eq:ETRH_ZETA} if we assume $E \propto GM^2/r_{\rm h}$ and we assume mass loss from the entire system is 
negligible. Since $t_{\rm rh}\propto r_{\rm h}^{\frac{3}{2}}$ it follows that $\displaystyle{r_{\rm h} \simeq r_{\rm h,i}\big(1+ \frac{3\zeta}{2t_{\rm rh,i}}(t-t_{\rm cc}) \big)^{\frac{3}{2}}}$, where powered expansion starts at time $t_{\rm cc}$ (the reason for this notation will become clear in Section \ref{sec:Nbody}). Therefore $\dot{M}_2$ can be expressed as

\begin{equation}\label{eq:torecalM2dot}
\hspace{1cm} \dot{M_2} \simeq -\frac{Cr_{\rm h,i}^{-\frac{3}{2}}}{1+\displaystyle{\frac{3\zeta}{2t_{\rm rh,i}}}(t-t_{\rm cc})}
\end{equation}
and if we integrate the above equation we get
\begin{equation}\label{eq:masslosslog}
  \; \; \; \; \; \; \;\; \;  \;\; \;   M_2 \simeq M_{\rm 2,i} - \frac{2}{3}\frac{\alpha}{\beta} M \ln\big({1+\displaystyle{\frac{3\zeta}{2t_{\rm rh,i}}} (t-t_{\rm cc})}\big).
\end{equation}

Throughout this section it has been assumed that $\zeta$ is a constant. However, as discussed in Section \ref{sec:Nbody}, $\zeta$ has been found to vary with time in situations where the BH sub-system cannot provide enough energy for balanced evolution. Even if $\zeta$ is time-dependent equation \ref{eq:M2dot_general} and the equation
$$\frac{1}{r_{\rm h}}\dot{r}_{\rm h} = \frac{1}{t_{\rm rh}} \zeta,$$ 
are still expected to hold (under the other assumptions made in this section). These two equations can be combined into a single equation which relates $\dot{M}_2$ to $\dot{r}_{\rm h}$ and has no explicit $\zeta$ dependence (i.e. $\displaystyle{\dot{M}_2 = -M\frac{\alpha}{\beta}\frac{1}{r_{\rm h}}\dot{r}_{\rm h}}$). The resulting equation can be easily solved (assuming that the variation of $M$ and $\alpha/\beta$ are neglected) and its solution is
\begin{equation}\label{eq:MvsRH}
\hspace{2cm}M_2 = M_{\rm 2,i} - M\frac{\alpha}{\beta}\ln{\frac{r_{\rm h}}{r_{\rm h,i}}}.
\end{equation}
This result implies that, regardless of $\zeta$, systems with the same $M_{\rm 2,i}$ and $r_{\rm h,i}$ should evolves along the same curve in  $M_{2}$, $r_{\rm h}$ space.

\subsection{Tidally limited systems}\label{sec:tidal}
In this section we will briefly consider the theory of tidally limited systems containing BH sub-systems. In H\'{e}non'’s tidally limited model \citep{Henon1961}, the rate of mass loss is 
\begin{equation}\label{eq:henontidal}
\hspace{32mm} \dot{M} = -\frac{M}{t_{\rm rh}}\xi
\end{equation}
where $\xi$ is a constant ($\xi=0.045$). In Section \ref{sec:ej} an equation of the same form was found for $\dot{M}_2$, i.e. 
$$ \dot{M}_2 =  -\frac{M}{t_{\rm rh}} \frac{ \alpha\zeta}{\beta}.$$
The relation between $\dot{M}$ and  $\dot{M_2}$ can be found by simply dividing these two equations, which results in 

$$ \frac{\dot{M}_2}{\dot{M}} = \frac{\alpha \zeta}{\beta \xi} = 0.11 \frac{\alpha}{0.15} \frac{\zeta}{0.0725}\frac{2.2}{\beta}  \frac{0.045}{\xi}  .$$
Therefore $\dot{M}_2/\dot{M}$ is a constant. For canonical values of the constants in the above equation $\dot{M}_2/\dot{M}\approx 0.11$. Note that the tidally limited model has a different value of $\zeta$ ($\zeta=0.0725$, see \citealt{Henon1961}) than for an isolated model ($\zeta=0.0926$, see \citealt{Henon}). The constant value of $\dot{M}_2/\dot{M}$ implies that for two-component systems there is a threshold value of $M_2/M_1$ at $\sim 10^{-1}$ above which $M_2/M$ (and hence $M_2/M_1$) is expected to grow with time and below which $M_2/M$ decreases with time. In other words if $M_2/M_1\gtrsim 10^{-1}$ then the system is expected to become more BH dominated, ultimately becoming a so-called dark star cluster \citep{Banerjee2011}. Alternatively if $M_2/M_1\lesssim 10^{-1}$ then the BH sub-system is expected to dissolve. In Section \ref{sec:classification}, where two-component parameter space is classified into different regions (see 
Fig. 
\ref{fig:gtoparameter}), there is already a distinction at roughly $M_2/M_1\sim10^{-1}$ (between region II \& III systems), based on other reasons discussed in that section. The theory in this section can be viewed as another reason for the distinction.

It is important to note that this  result has yet to be rigorously tested  because tidally limited systems are not considered further in this paper and the exact threshold value is likely to depend on a number of astrophysical issues (e.g. initial mass function, tidal shocks etc). Indeed while equation \ref{eq:henontidal} is a reasonable approximation, \cite{Baumgardt2001} showed that the time scale of escape depends on both $t_{\rm rh}$ and the crossing time. Nevertheless it seem likely that a threshold value of $M_2/M_1$ exists even for more realistic systems, although it may have some dependence on the other properties of the system (e.g. $m_2/m_1$).

\section{Dependence of \lowercase{$r_{\rm h,2}/r_{\rm h}$} on cluster parameters}\label{sec:rhsec}
\subsection{Gas models}\label{sec:rhsecgas}
\begin{figure}
\subfigure{\scalebox{0.65}{\includegraphics{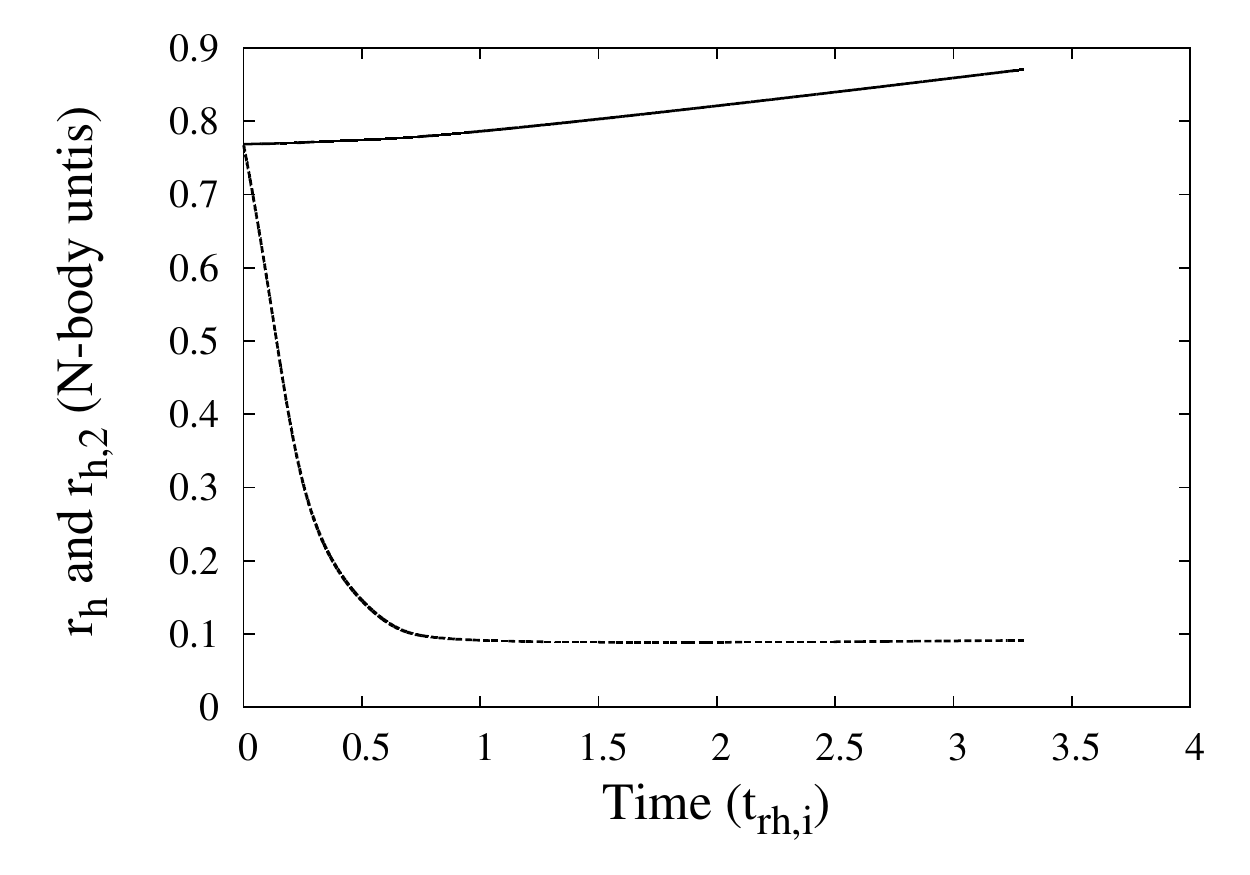}}}\quad
\subfigure{\scalebox{0.65}{\includegraphics{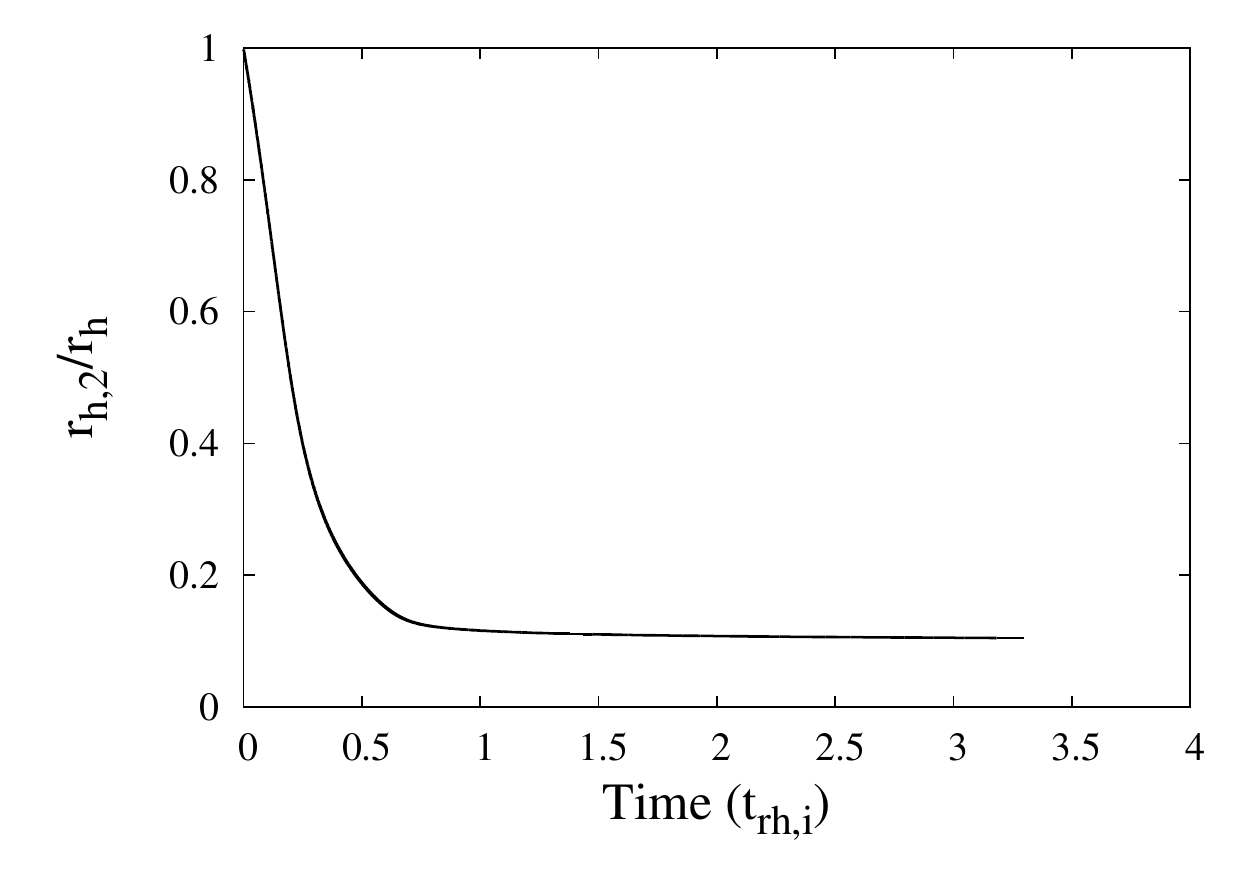}}}\quad
\caption{ Top: $r_{\rm h}$ (top line) and $r_{\rm h,2}$ (bottom line) vs time (units $t_{\rm rh,i}$) of gas models with $N=32k$, $m_2/m_1=10$ and $M_2/M_1=0.02$. $r_{\rm h}$ and $r_{\rm h,2}$ are given in N-body units. Initially $r_{\rm h}$ and $r_{\rm h,2}$ have the same value, but mass segregation quickly decreases $r_{\rm h,2}$, after which it reaches an approximately steady value. Bottom: $r_{\rm h,2}/r_{\rm h}$ vs time (units $t_{\rm rh,i}$). Core collapse (see Section \ref{sec:theory}) occurs at $\approx 1t_{\rm rh,i}$; shortly before this $r_{\rm h,2}/r_{\rm h}$ reaches a nearly constant value.}
\label{fig:rh1}
\end{figure}

The aims of Sections \ref{sec:rhsecgas} and \ref{sec:rhsecnbody} are to test the dependence 
of $r_{\rm h,2}/r_{\rm h}$ on the cluster parameters and compare the results 
with the theory presented in Section \ref{sec:theory}. The simulations in 
this section (\ref{sec:rhsecgas}) were run using a two-component gas code
 (see \cite{HeffieAarseth1992} and \cite{breenheggie1}). In all cases, the initial conditions used were realisations of the Plummer model \citep{Plummer1911, HeggieHut2003}. The initial velocity dispersion of both components and the initial ratio of density of both components were equal at all locations.  The choices for the ratio of stellar masses  ($m_{2}/m_1$) are $2$, $5$, $10$, $20$, $50$ and $100$. The values of the total mass ratio ($M_{2}/M_1$) used in this section are $0.5$, $0.1$, $0.05$, $0.02$ and $0.01$, though the first two values are outside the parameter space of interest in most of the present paper. The value of $r_{2}/r_{\rm h}$, which was found to be approximately constant 
during 
the post collapse phase of evolution (see Fig. \ref{fig:rh1}), was measured for the series of models and is given in Table \ref{table:rBHrh}.

\begin{table}
\begin{center}
\caption{Values of $r_{\rm h,2}/r_{\rm h}$ in post collapse evolution ($N=32k$). These values where measured over $1t_{\rm rh,i}$ after a time of at least $2t_{\rm cc}$, where $t_{\rm cc}$ is the time of core collapse. For low $M_2/M_1$ ($<0.1$) there is a clear trend of increasing $r_{\rm h,2}/r_{\rm h}$ with increasing $m_2/m_1$. The total mass ratios of $0.5$ and $0.1$ have also been included to demonstrate that there is an inverse dependence of $r_{\rm h,2}/r_{\rm h}$ on $m_2/m_1$ for the largest value of $M_2/M_1$ considered. See text for more details.}
\begin{tabular}{ c  |c |c| c|c |c |}
 \bf{$\frac{m_{2}}{m_1}\backslash\frac{M_{2}}{M_1}$}   &  \bf{0.5}   & \bf{0.1}   & \bf{0.05}   & \bf{0.02}& \bf{0.01}     \\ \cline{1-6} 
 \bf{100}        & 0.37  & 0.34  & 0.29 & 0.24 & 0.21 \\ 
 \bf{50}         & 0.38  & 0.27  & 0.24 & 0.19 & 0.18 \\ 
 \bf{20}         & 0.40  & 0.21  & 0.18 & 0.15 & 0.13 \\ 
 \bf{10}         & 0.42  & 0.18  & 0.14 & 0.11 & 0.09 \\ 
 \bf{5}          & 0.44  & 0.18  & 0.11 & 0.07 & 0.04 \\ 
 \bf{2}          & 0.50  & 0.20  & 0.10 & 0.03 & 0.02 \\ \
\label{table:rBHrh}
\end{tabular}
\end{center}
\end{table}

\begin{figure}
\subfigure{\scalebox{0.60}{\includegraphics{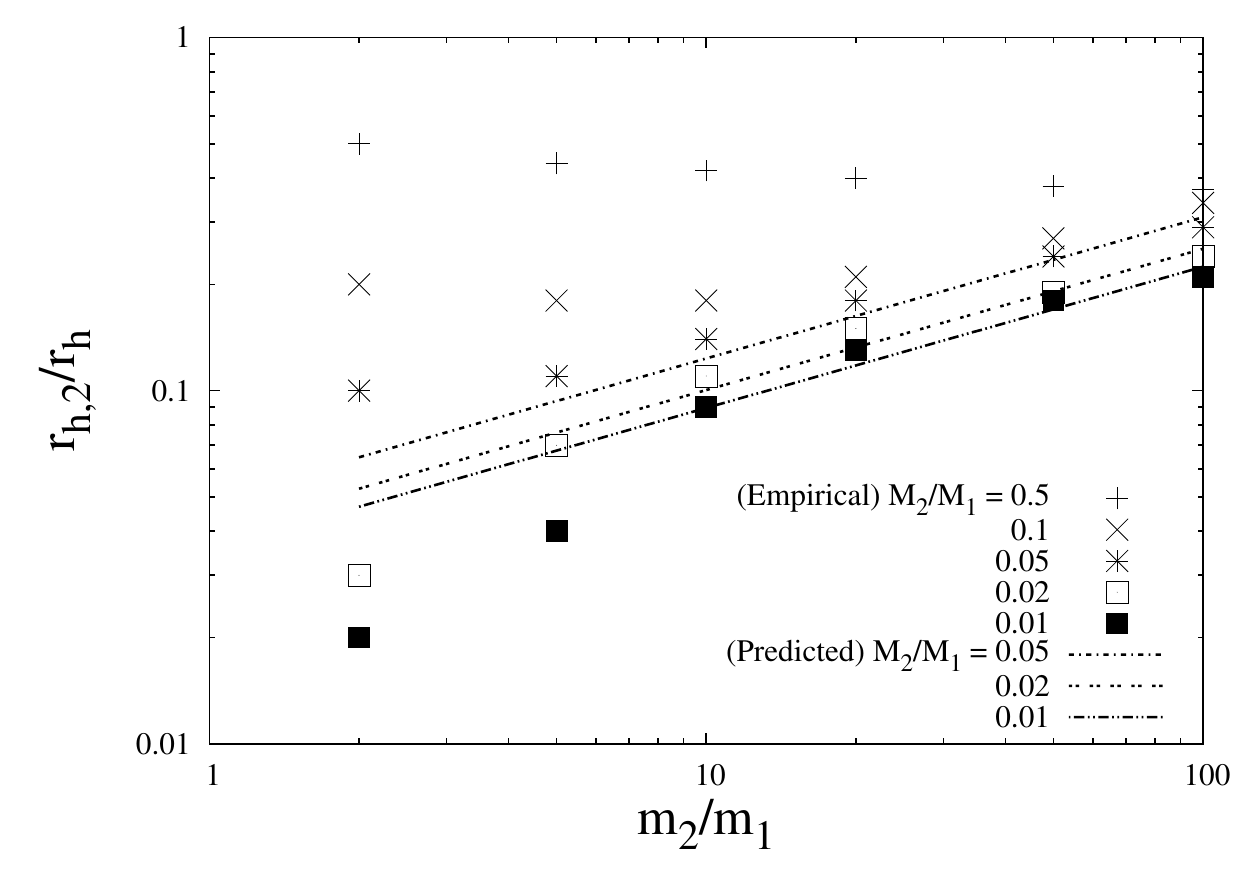}}}\quad
\caption{The variation of $r_{\rm h,2}/r_{\rm h}$ with $m_2/m_1$. The points represent the values of $r_{\rm h,2}/r_{\rm h}$ given in Table \ref{table:rBHrh}. The lines represent the expected variation of $r_{\rm h,2}/r_{\rm h}$ with $m_2/m_1$ (fitted curves of the form $b(m_2/m_1)^{0.4}$) as given by the theory in Section \ref{sec:theory}. The lines have only been included for $M_2/M_1 < 0.1$ as this is where the theory is expected to apply. The empirical measured variation of $r_{\rm h,2}/r_{\rm h}$ with $m_2/m_1$ is in good agreement with theory in Section  \ref{sec:theory}, when $m_2/m_1 \gtrsim  10$ and $M_2/M_1 < 0.1$. See text for further details.}
\label{fig:rBHrh2vari}
\end{figure}

The results in Table \ref{table:rBHrh} are plotted in Fig. \ref{fig:rBHrh2vari}. As can be seen in Fig. \ref{fig:rBHrh2vari} (and Table \ref{table:rBHrh}) the results are in qualitative agreement with the theory in Section \ref{sec:theory}, in the sense that for $M_2/M_1<0.1$ there is an increase in the values of $r_{\rm h,2}/r_{\rm h}$ with increasing $m_2/m_1$. For $M_2/M_1=0.5$ the trend is qualitatively different than for $M_2/M_1<0.1$; there is an increase in the values of $r_{\rm h,2}/r_{\rm h}$ with decreasing $m_2/m_1$. The theory in Section \ref{sec:theory} cannot be expected to apply in this regime, as the light component does not dominate. In this regime the decrease of  $r_{\rm h,2}/r_{\rm h}$ with $m_2/m_1$  can be explained qualitatively by the fact that if the BH have larger stellar masses then there is a stronger tendency towards mass segregation (see \cite{breenheggie1} for a discussion of this topic).

We now consider the comparison with theory more quantitatively in the regime $M_2/M_1 \lesssim 0.1$. In Fig. \ref{fig:rBHrh2vari} we can see that values of $r_{\rm h,2}/r_{\rm h}$ are also in \emph{quantitative} agreement with equation \ref{eq:fin} for $m_2/m_1 \gtrsim  10$ in the sense that the power law index is approximately confirmed. However $r_{\rm h,2}/r_{\rm h}$ increases more rapidly then is expected by equation \ref{eq:fin} for $m_2/m_1 < 10$ and $M_2/M_1 < 0.05$.  This behaviour is possibly explained by equation \ref{eq:balonestable}, which predicts a different power law index for Spitzer stable systems than in equation \ref{eq:fin}. Indeed for the lowest values of $M_2/M_1$ the slope is approximately consistent with equation \ref{eq:balonestable}. Nevertheless the stellar mass ratios of interest are $m_2/m_1 \gtrsim  10$ as realistic ratios for systems containing BH sub-systems would be in this range.


\begin{figure}
\subfigure{\scalebox{0.65}{\includegraphics{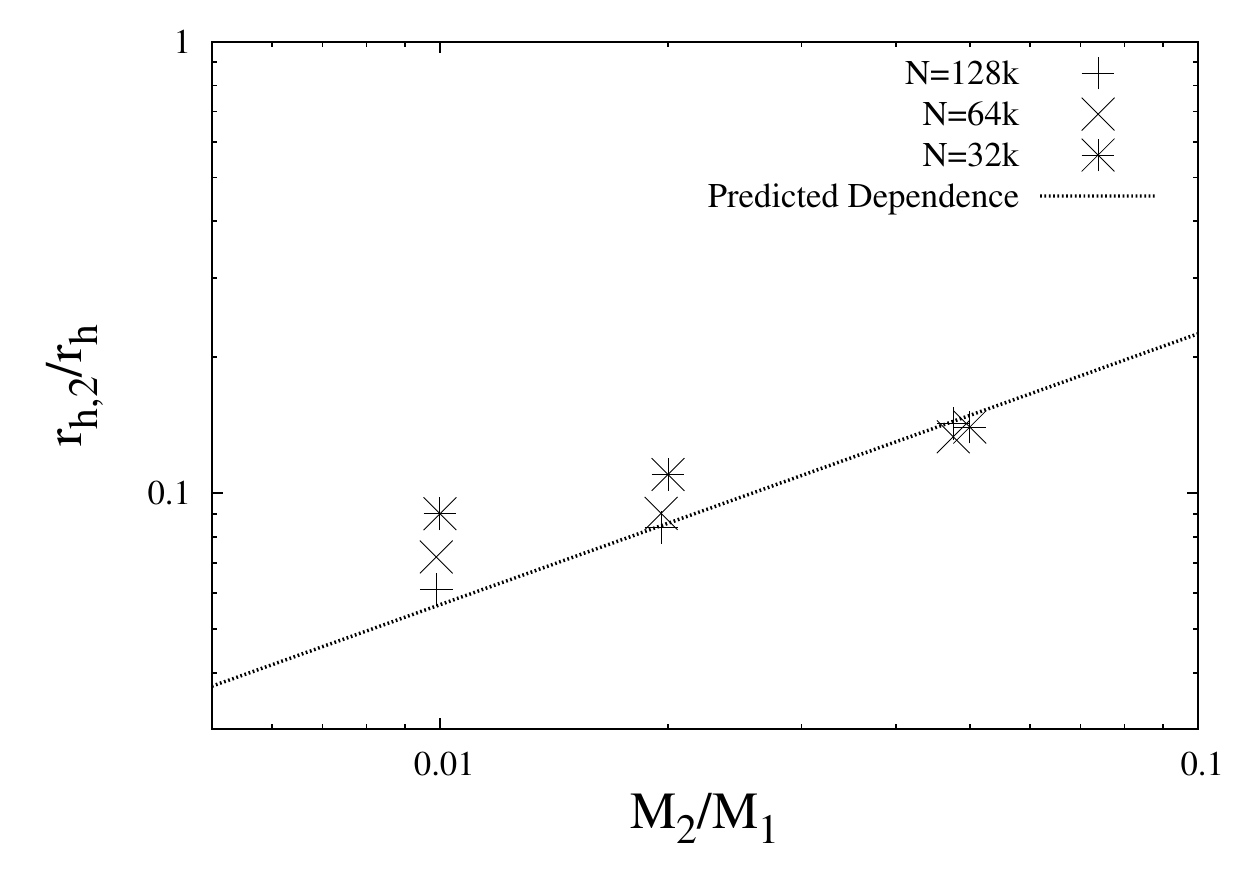}}}\quad
\caption{The variation of $r_{\rm h,2}/r_{\rm h}$ with $M_2/M_1$. The points represent the values of $r_{\rm h,2}/r_{\rm h}$ for the case $m_2/m_1=10$ with $N=32k$, $N=64k$ and $N=128k$. The line represents the expected variation of $r_{\rm h,2}/r_{\rm h}$ with $M_2/M_1$ as given by the theory in Section \ref{sec:theory}.}
\label{fig:rBHrh2variM}
\end{figure}

The variation of $r_{\rm h,2}/r_{\rm h}$ with $M_2/M_1$ over a range of different $N$ is shown in Fig. \ref{fig:rBHrh2variM}, for the case of $m_2/m_1=10$. For the case of $N=32k$ the variation is less than expected from equation \ref{eq:fin} (see Section  \ref{sec:theory}). The variation is approximately of the same form, i.e.  $r_{\rm h,2}/r_{\rm h}  \propto (M_2/M_1)^{a}$, but $a \approx 0.3$ for the case of $m_2/m_1=10$ and $N=32K$, which is less than the expected value of $a \simeq  0.6$. (Here we ignore the dependence on the Coulomb logarithms). The variation of $r_{\rm h,2}/r_{\rm h}$ with $M_2/M_1$  comes into better agreement with equation \ref{eq:fin} with increasing $N$, with the case of $N=128k$ being in good agreement with the theory in Section  \ref{sec:theory}. This seems to indicate that the disagreement is caused by small values of $N_2$. One of the assumptions under which equation \ref{eq:fin} is 
derived is that  $r_{\rm c} \ll r_{\rm h,2}$, but it is 
possible that  
$r_{\rm c,2} \approx r_{\rm h,2}$ for small $N_2$ as shown in equation \ref{eq:balone}. In this case one of the assumptions  underlying the theory is not satisfied.

\subsection{$r_{\rm h,2}/r_{\rm h}$ in N-body runs}\label{sec:rhsecnbody}

\begin{figure}
\subfigure{\scalebox{0.7}{\includegraphics{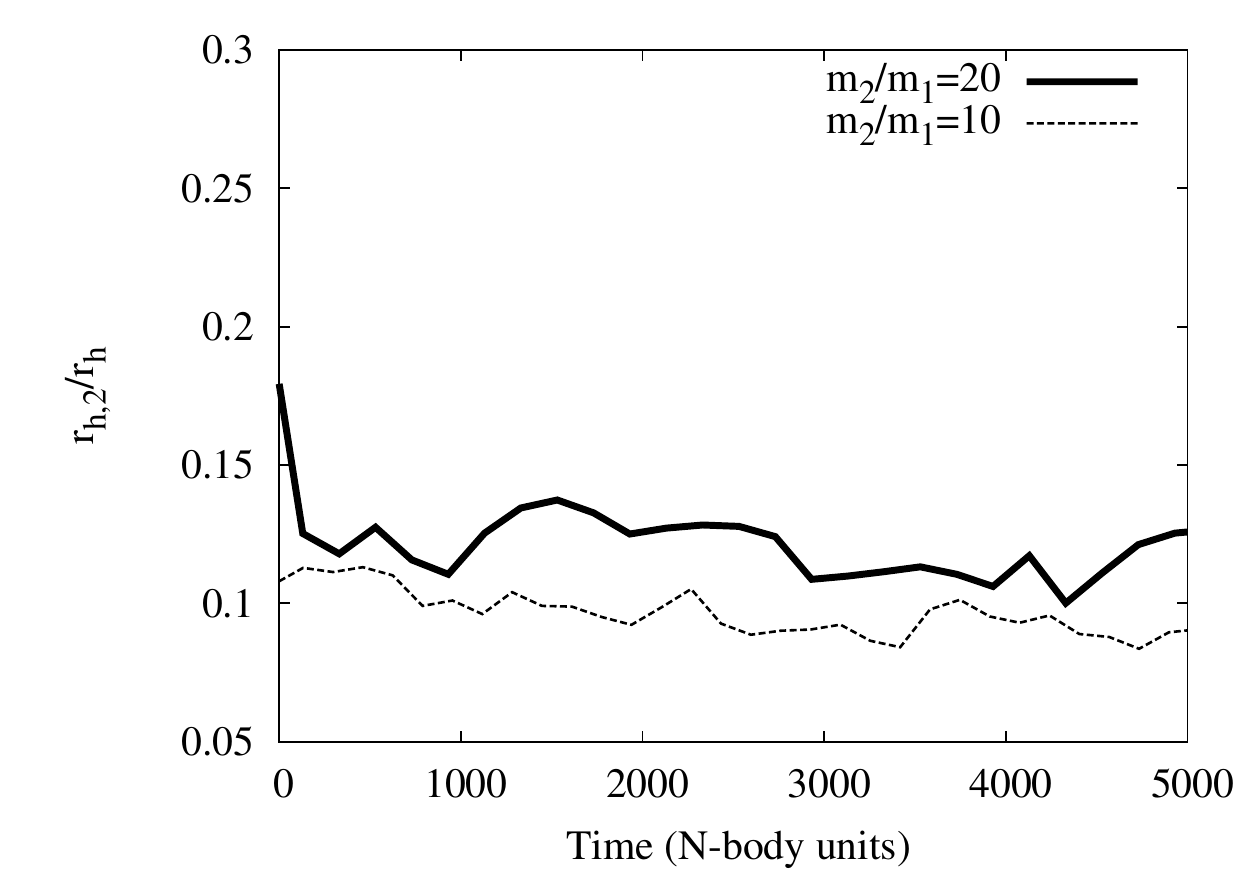}}}\quad
\caption{$r_{\rm h,2}/r_{\rm h}$ vs time (in N-body units). N-body runs with initial values   $N=64k$, $M_2/M_1=0.02$, $m_2/m_1=20$ (thick line) and $m_2/m_1=10$ (thin line). Time is set so that core collapse occurs at $t=0$ for both systems. $r_{\rm h,2}/r_{\rm h}$ has been smoothed to make the plot clearer. The values of $r_{\rm h,2}/r_{\rm h}$ in the graph are in approximate agreement with the results from the two-component gas model given in Table \ref{table:rBHrh}.}
\label{fig:64krh2rh}
\end{figure}

In Section \ref{sec:theory} it was predicted that $r_{\rm h,2}/r_{\rm h}$ would increase 
with increasing $m_2/m_1$ in a system with fixed $M_{2}/M_1$ and $N$. This prediction 
will now be tested with direct N-body runs (see Section
 \ref{sec:Nbody}). The initial conditions are realisations of the Plummer Model with $N=64k$, $M_2/M_1=0.02$ and two different values of $m_2/m_1$ ($10$ and $20$). We have compared the values of $r_{\rm h,2}/r_{\rm h}$ in Fig. \ref{fig:64krh2rh}. The value of $r_{\rm h,2}/r_{\rm h}$ is indeed larger for $m_2/m_1=20$ than for $m_2/m_1=10$ as expected. This effect was confirmed using a two-component gas code in the previous subsection. However mass is conserved in the gas models whereas in the more realistic N-body systems mass is lost over time. Therefore we need to ensure that we are comparing the values of $r_{\rm h,2}/r_{\rm h}$ for both the runs at constant $M_2/M_1$. 
For the two runs in Fig. \ref{fig:64krh2rh} mass is lost from the BH sub-system 
at approximately the same rate (see Fig.\ref{fig:64kmassloss}, bottom), as predicted by the theory in Section \ref{sec:ej}. The mass loss from the light component is negligible over the length of these N-body runs: less than $5\%$ of the light component is lost during the entire run. Therefore while $M_2/M_1$ does decrease with time over the runs in Fig. \ref{fig:64krh2rh}, the values of $M_2/M_1$ for both runs are approximately the same at any given time. 

\begin{figure}
\subfigure{\scalebox{0.65}{\includegraphics{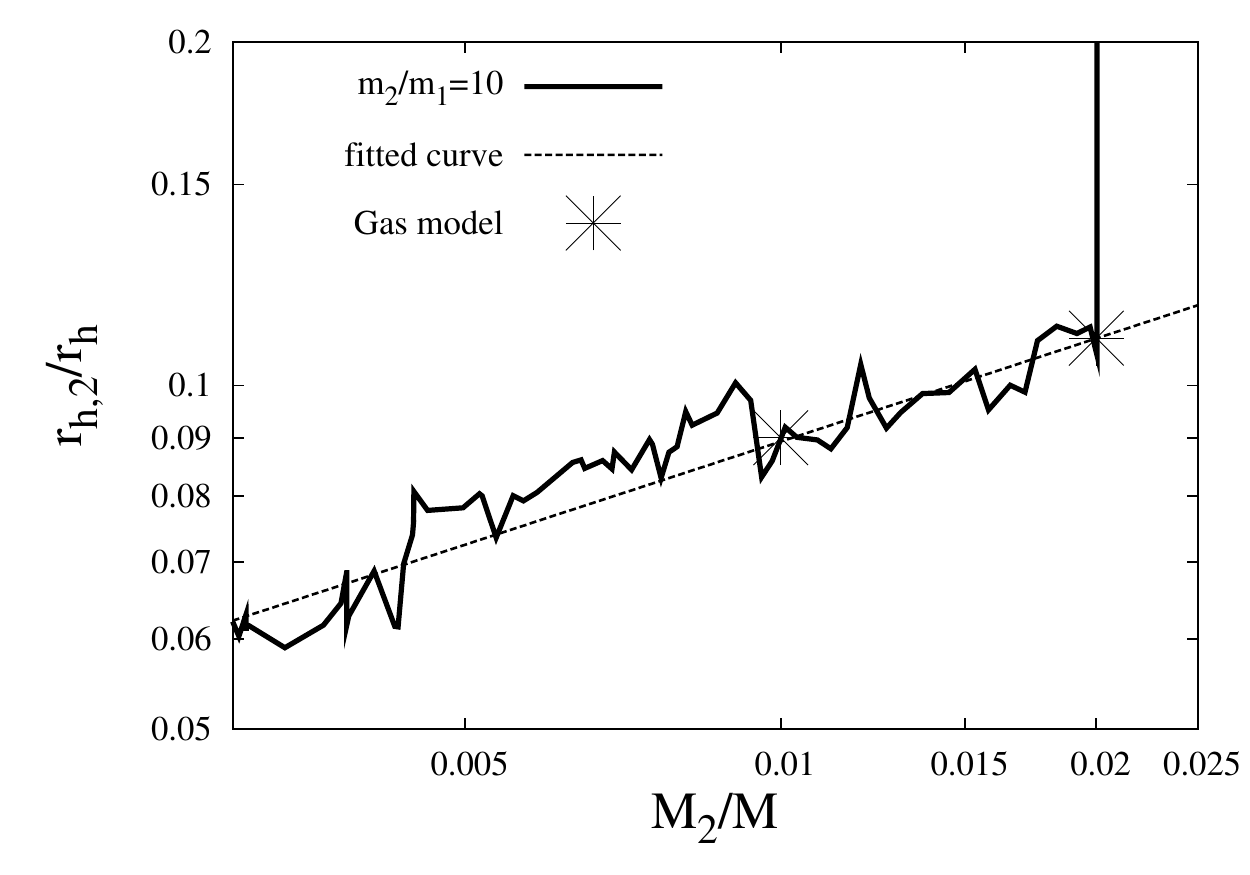}}}\quad
\caption{$r_{\rm h,2}/r_{\rm h}$ vs $M_2/M_1$. The solid line represents results from an N-body run with initial values $N=64k$, $M_2/M_1=0.02$ and $m_2/m_1=10$. The results are smoothed to make the value of $r_{\rm h,2}/r_{\rm h}$ clearer. The dotted line is the best matching curve of the form $b\big(M_2/M_1)^a$ (where $a$ and $b$ are constants, the best match values being $b\approx 0.33$ and $a\approx 0.28$). The points are results from the two-component gas model (see Table \ref{table:rBHrh}), where $M_2/M_1$ is fixed. At the beginning of the run $r_{\rm h,2}/r_{\rm h}=1$ before mass segregation rapidly reduces its value, whence the vertical line segment in the top right corner.}
\label{fig:64krh2rh22}
\end{figure}

The variation of $r_{\rm h,2}/r_{\rm h}$ with $M_2/M_1$ is shown in Fig. \ref{fig:64krh2rh22} for the N-body run with $N=64k$, $m_2/m_1=10$ and $M_2/M_1=0.02$. Initially $r_{\rm h,2}/r_{\rm h}=1$ before being quickly reduced due to mass segregation. The BH sub-system starts producing energy when  $r_{\rm h,2}/r_{\rm h}$ reaches $\approx 0.1$ and $M_2/M_1$ begins to decrease at this point. The variation of $r_{\rm h,2}/r_{\rm h}$ with $M_2/M_1$ is again less then expected from equation \ref{eq:fin}, with the result indicating a dependence of  $r_{\rm h,2}/r_{\rm h} \propto (M_2/M_1)^{0.28}$. Although this is not in agreement with equation \ref{eq:fin}  if we neglect the coulomb logarithm, the results are in good agreement with the gas model. 


\subsection{Central potential}\label{sec:phi}
\begin{figure}
\subfigure{\scalebox{0.70}{\includegraphics{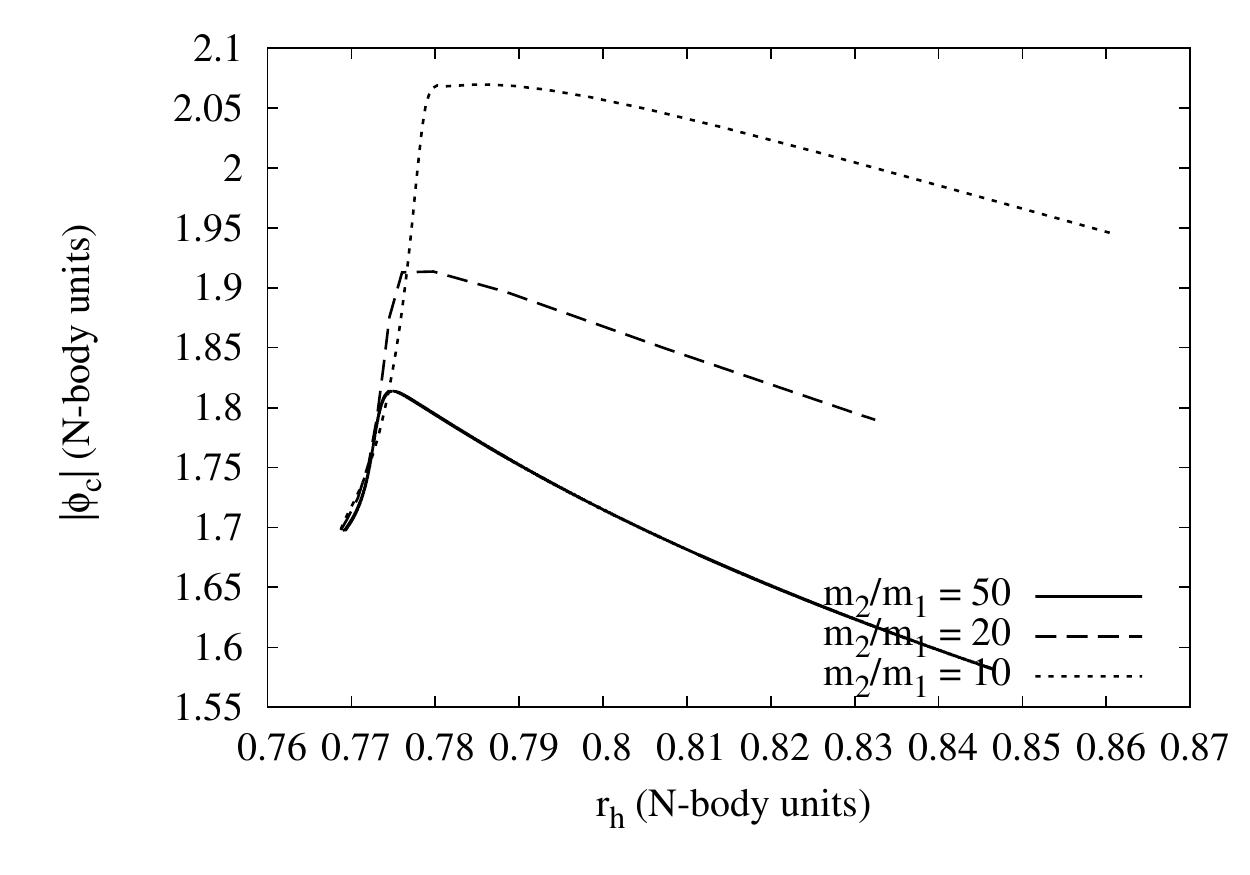}}}\quad
\subfigure{\scalebox{0.70}{\includegraphics{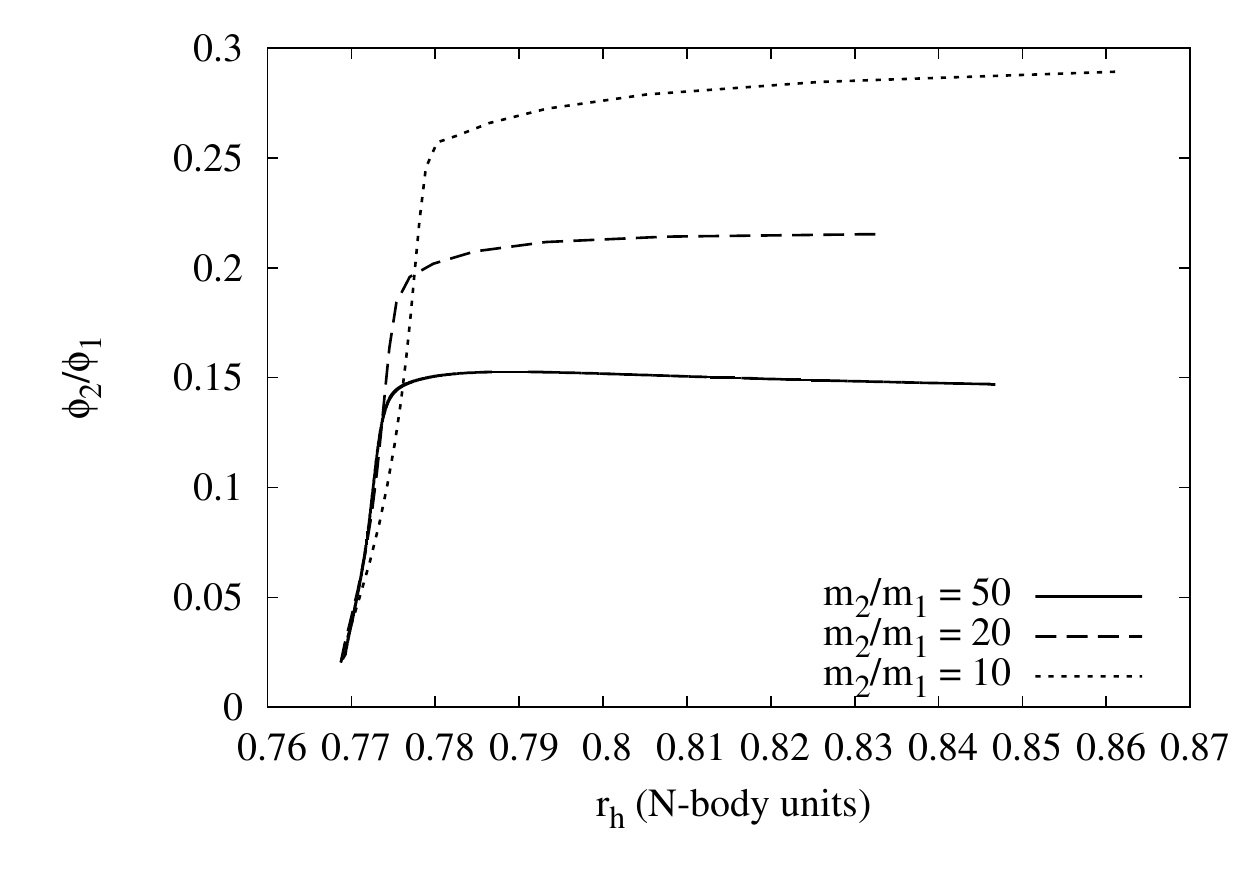}}}\quad
\caption{ Top: $|\phi_{\rm c}|$ vs $r_{\rm h}$ for gaseous systems with fixed $N$ ($128k$) and $M_2/M_1=0.02$; the different values of $m_2/m_1$ are $10$, $20$ and $50$. Bottom: $\phi_2/\phi_1$ vs $r_{\rm h}$ for the same systems as in the top figure.  This plot shows that the contribution of the light component to the central potential is dominant.}
\label{fig:phi1}
\end{figure}

In section \ref{sec:ej} we made the assumption that the main contribution to the central potential is from the light component. In order to test this assumption the central potential and the relative contribution of each component to the central potential ($\phi_2/\phi_1$), have been measured in a series of two-component gas models with $M_2/M_1=0.02$. These results are presented in Table \ref{table:phit}. The values in Table \ref{table:phit} were measured once the systems had reached a certain value of $r_{\rm h}$ ($r_{\rm h}=0.83$). This was done in order to insure the systems had reached a similar point in their evolution (see Fig. \ref{fig:phi1}). The variation in $\phi_{\rm c}$ in Table \ref{table:phit} is only about a factor of $1.2$ for fixed $N$ even though $m_2/m_1$ varies by a factor of $5$. The variation in $\phi_2/\phi_1$ is higher, but the values are of the same order of magnitude as the estimate in Section \ref{sec:ej} and indeed in satisfactory agreement, considering that 
$M_2/M_1$ is higher here. As the energy generation rate per unit mass is $\propto m_2^3\rho_2^2/\sigma_2^7$ \citep{HeggieHut2003}, for systems with the same value of $N$, $M_2/M_1$ and $m_1$, as $m_2$ increases the system can produce the required energy at a lower central density, thus the central potential is expected to becomes shallower with increasing $m_2/m_1$ as seen in Table \ref{table:phit}. This is also why two-component systems with larger $m_2/m_1$ are stable against gravothermal oscillation to higher values of $N$ then for lower $m_2/m_1$  \citep{breenheggie1}.

These values serve as a rough guide to how the variation of $m_2/m_1$ will affect the system. If we consider systems with fixed $N$, $M_2$ and $M_2/M_1$ and use similar reasoning as in Section \ref{sec:ej}, we would expect systems with lower $m_2/m_1$ to last slightly longer than systems with higher $m_2/m_1$. This is because the average energy contribution per binary is dependent on the depth of the central potential. The deeper the central potential the greater the average energy contribution per binary will be, and therefore it is expected that for fixed $M_2/M_1$ a two-component system with $m_2/m_1=20$ will lose mass slightly faster than a system with $m_2/m_1=10$.

\begin{table}
\begin{center}
\caption{Variation of $|\phi_{\rm c}|$ and $\phi_2/\phi_1$ with $m_2/m_1$ for systems with $M_2/M_1=0.02$. The values are measured in the post-collapse phase of evolution when the systems have reached a certain size ($r_{\rm h} \approx 0.83$).}
\begin{tabular}{ c  |c |c|c| c|}\label{table:phit}
\bf{$m_{2}/m_1$}&\multicolumn{2}{|c|}{$N=32k$}  & \multicolumn{2}{|c|}{$N=128k$}  \\ \cline{1-5}
 &   $|\phi_{\rm c}|$ & $\phi_2/\phi_1$  & $|\phi_{\rm c}|$ & $\phi_2/\phi_1$     \\ 
\bf{50}     & 1.55 & 0.10 &  1.62 & 0.15 \\ 
\bf{20}     & 1.67 & 0.14 &  1.80 & 0.22 \\ 
\bf{10}     & 1.85 & 0.18 &  2.00 & 0.28    
\end{tabular}
\end{center}
\end{table}

\section{Evolution of the BH sub-system: Direct N-body Simulations}\label{sec:Nbody}
\subsection{Overview}\label{sec:nbody_intro}

\begin{figure}
\subfigure{\scalebox{0.62}{\includegraphics{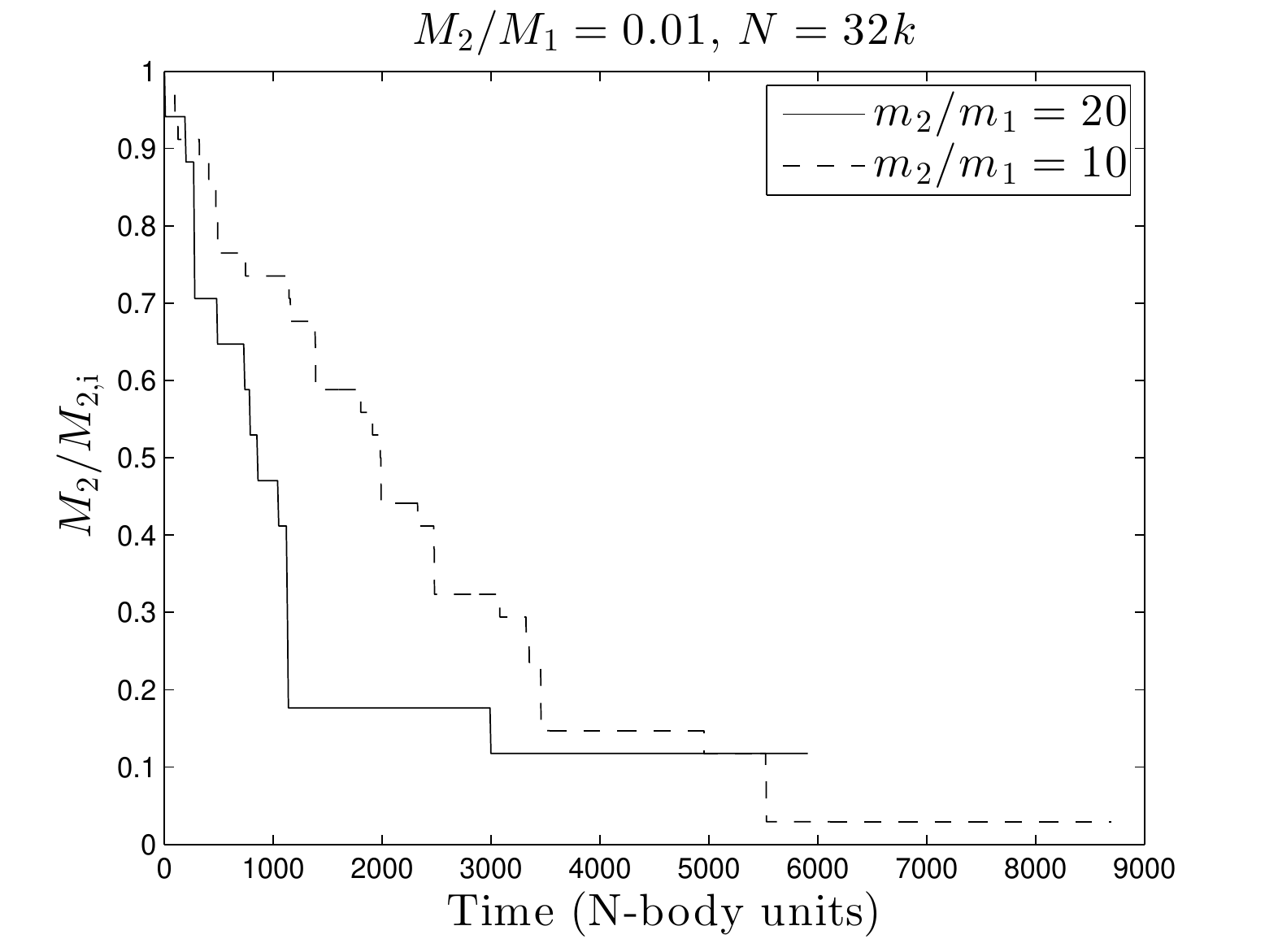}}}\quad
\subfigure{\scalebox{0.62}{\includegraphics{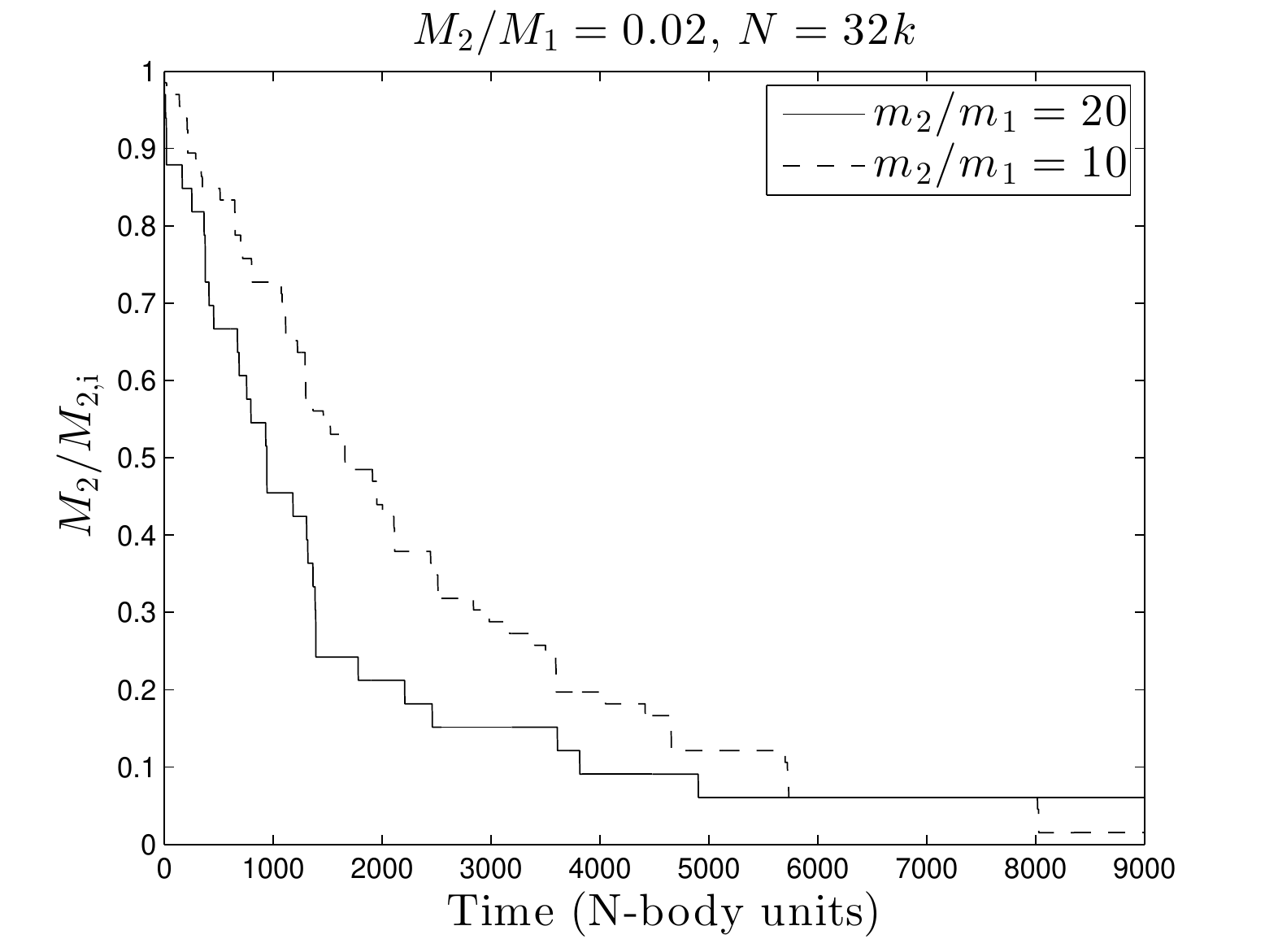}}}\quad
\caption{Fraction of initial mass remaining ($M_2/M_{\rm 2,i}$, where $M_{\rm 2,i}$ is the initial mass of the heavy component) vs time (in N-body units) for the cases $N=32k$ with $M_2/M_1=0.01$ (Top) and  $M_2/M_1=0.02$ (Bottom). In both figures the dashed line represents $m_2/m_1=10$ and the solid line represents $m_2/m_1=20$. $T=0$ is set as the time when first mass loss occurs, which is at approximately the same time as the first core collapse.}
\label{fig:32kM01massloss}
\end{figure}

\begin{figure}
\subfigure{\scalebox{0.62}{\includegraphics{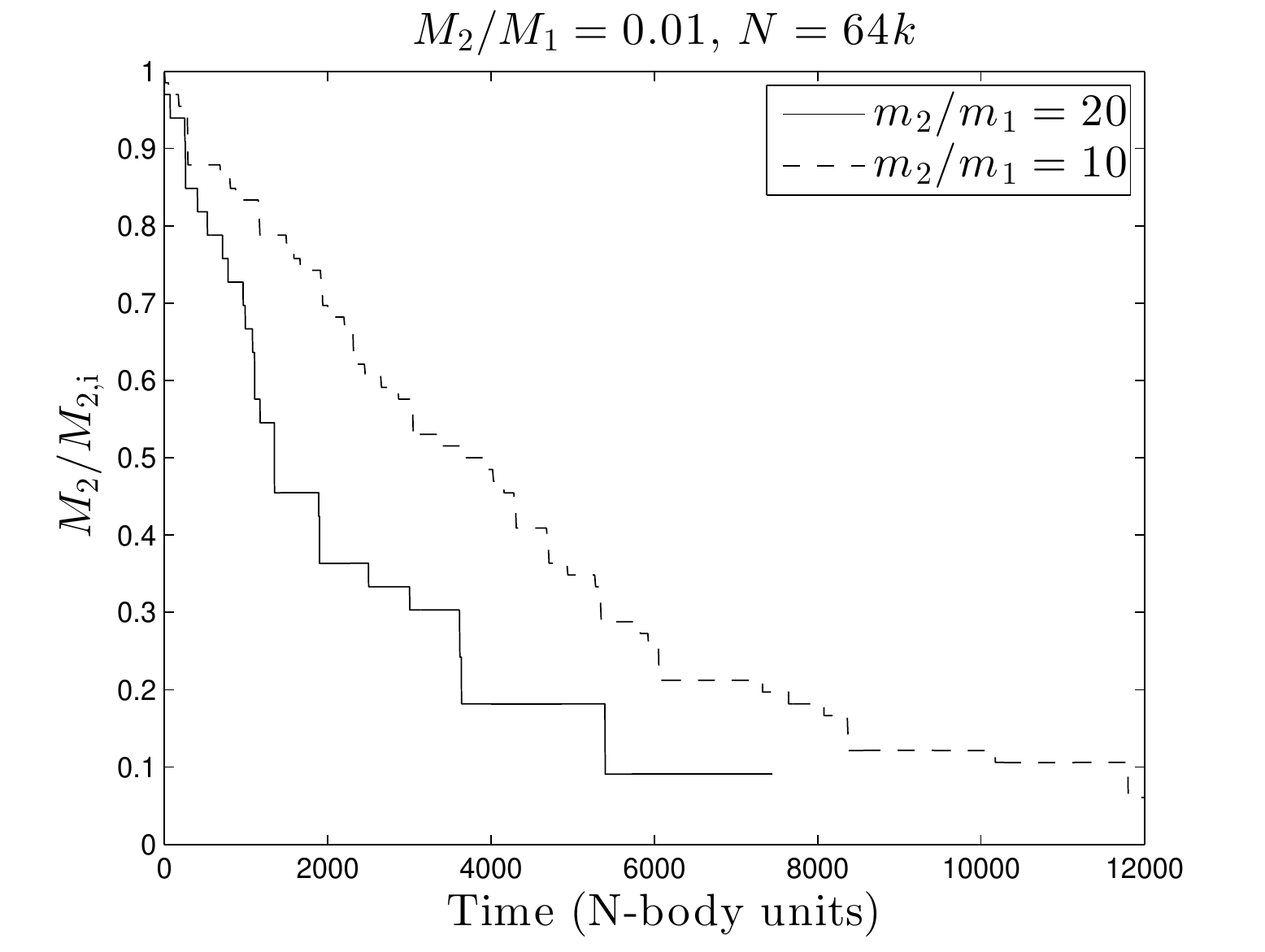}}}\quad
\subfigure{\scalebox{0.62}{\includegraphics{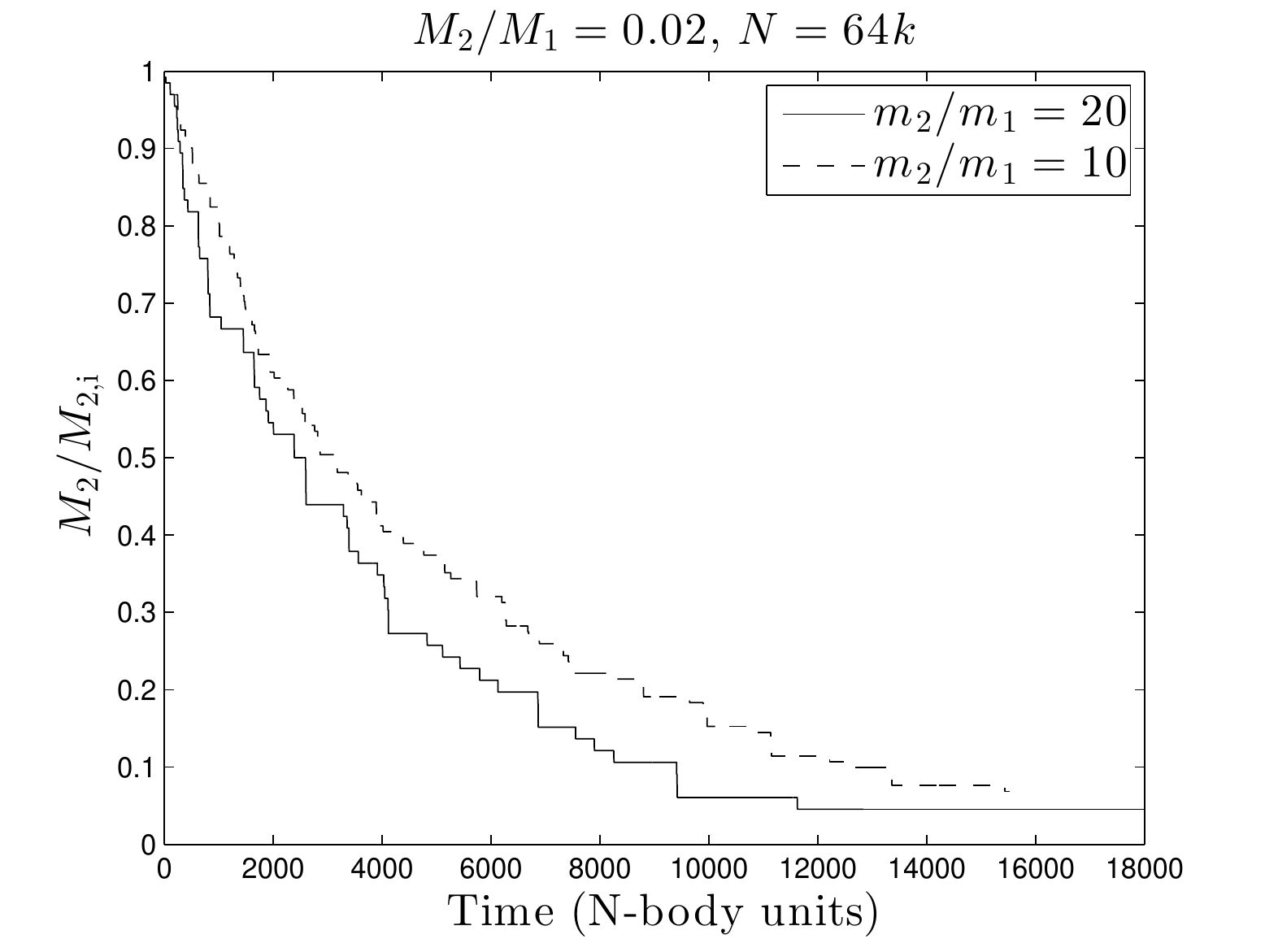}}}\quad
\caption{Fraction of initial mass remaining ($M_2/M_{\rm 2,i}$, where $M_{\rm 2,i}$ is the initial mass of the heavy component) vs time (in N-body units) for the cases $N=64k$ with $M_2/M_1=0.01$ (Top) and  $M_2/M_1=0.02$ (Bottom). In both figures the dashed line represents $m_2/m_1=10$ and the solid line represents $m_2/m_1=20$. $T=0$ is set at the time when the first mass loss occurs, which is at approximately the same time as the first core collapse.}
\label{fig:64kmassloss}
\end{figure}

\begin{figure}
\subfigure{\scalebox{0.60}{\includegraphics{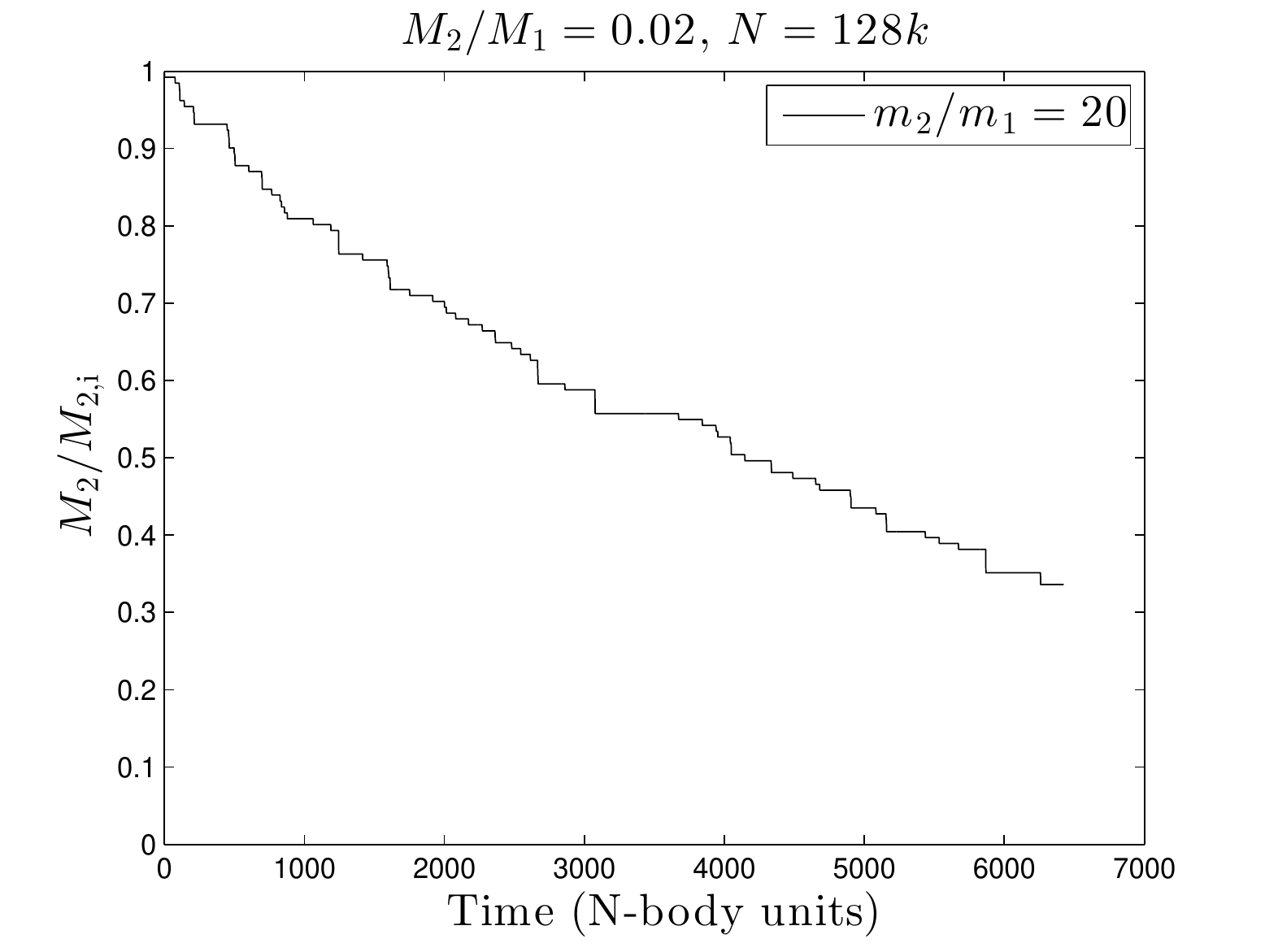}}}\quad
\caption{Fraction of initial mass remaining ($M_2/M_{\rm 2,i}$, where $M_{\rm 2,i}$ is the initial mass of the heavy component) vs time (in N-body units) for the case $N=128k$ with $M_2/M_1=0.02$ and  $m_2/m_1=20$. $T=0$ is set at the time when the first mass loss occurs, which is at approximately the same time as the first core collapse.}
\label{fig:128kmassloss}
\end{figure}

In order to study BH sub-systems we carried out a number 
of N-body simulations using the NBODY6 code \citep{NitadoriAaresth2012}. 
Because of the computational cost of large $N$ 
simulations we have mostly limited ourselves to runs with $N=32k$ 
and $64k$. The total mass ratios used were $M_2/M_1=0.01$ and $0.02$. The stellar mass ratios used were $m_2/m_1=10$ and $20$. An additional run with $N=128k$, $M_2/M_1=0.02$ and $m_2/m_1=20$ was also carried out to increase the number of systems with $N_2 > 10^2$. These parameters are summarised in Table \ref{table:Ndetails} and the results of these runs are given in Table \ref{table:lifeN1}. The evolution of the fraction of mass remaining in BH ($M_2/M_{\rm 2,i}$) in all runs is shown in Figs \ref{fig:32kM01massloss}, \ref{fig:64kmassloss} and \ref{fig:128kmassloss}.

\begin{table}
\begin{center}
\caption{Parameters used for N-body runs. In all cases the initial conditions were realisations of the Plummer Model \citep{Plummer1911}. The values of $t_{\rm rh,i}$ (the initial half-mass relaxation time) are calculated using $\Lambda=0.02N$ for the coulomb logarithm.}
\begin{tabular}{ c  |c |c|c|}
$N$     & $m_{2}/m_1$ & $M_{2}/M_1$ &  $t_{\rm rh,i}$ (N-body units)    \\ \cline{1-4} 
$32k$   &   $10,$ $20$             &  $0.01,$  $0.02$    &  $471$   \\
$64k$   &   $10,$ $20$             &  $0.01,$  $0.02$    &  $851$   \\
$128k$  &   $20$                   &   $0.02$    &  $1552$   
\label{table:Ndetails}
\end{tabular}
\end{center}
\end{table}
\begin{figure}
\subfigure{\scalebox{0.7}{\includegraphics{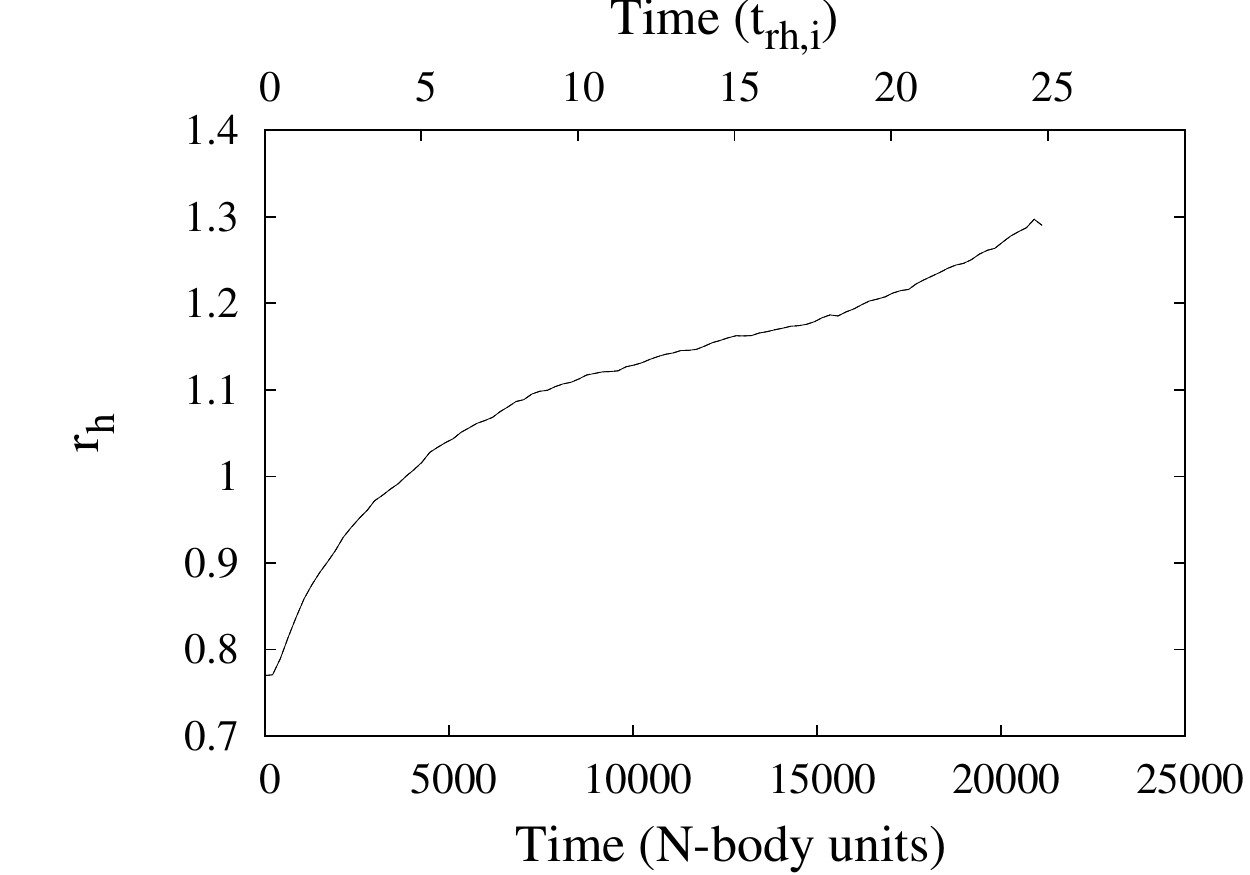}}}\quad
\subfigure{\scalebox{0.7}{\includegraphics{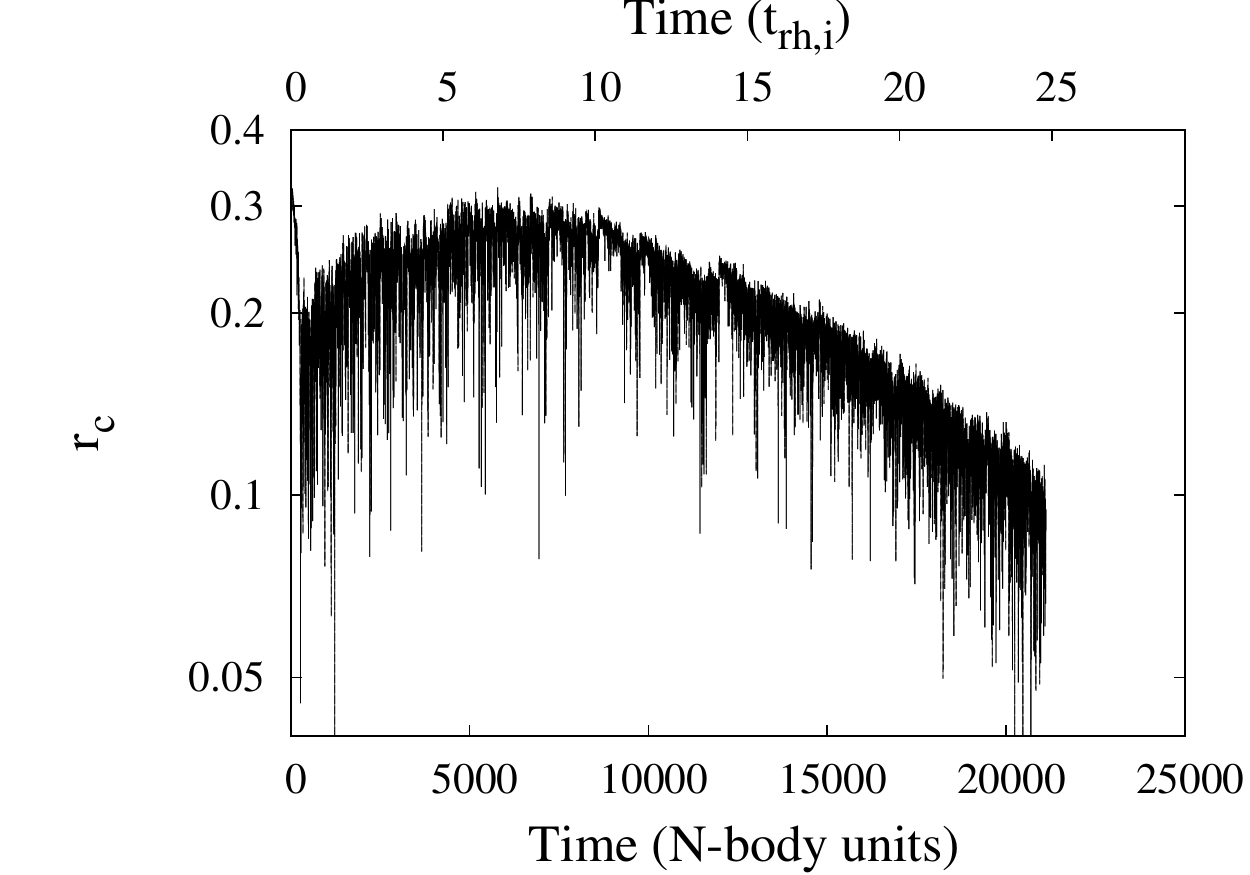}}}\quad
\caption{N-body run ($20$,$0.02$,$64k$), i.e. with $N=64k$, $m_2/m_1=20$ and $M_2/M_1=0.02$. Bottom: $r_{\rm c}$ vs time (N-body units). At $t=270$ N-body units core collapse occurs. After most of the BH have escaped ($t\approx 7500$) the core starts to re-collapse. Top: $r_{\rm h}$ vs time (N-body units). $r_{\rm h}$ initially expands rapidly before gradually slowing down as the BH sub-system dissolves. From $t\approx 10000$ to $\approx 15000$ there is little change in $r_{\rm h}$ as the system is no longer in a balanced energy generating phase of evolution. The expansion after  $t\approx 15000$ results from a single remaining BH binary which becomes more active as the core collapses.}
\label{fig:BEHA}
\end{figure}

First we will discuss the qualitative behaviour of systems 
containing a BH sub-system. We do this by considering
 the case $m_2/m_1=20$,  $M_{2}/M_1 =0.02$ and $N=64k$, which we 
refer to henceforth as ($20$,$0.02$,$64k$). The graphs of $r_{\rm h}$ and $r_{\rm c}$ against 
time for this run are shown in Fig. \ref{fig:BEHA}. The BH population 
quickly segregates to the centre of the system causing core 
collapse to occur in the BH sub-system. For the parameters of this model, this takes 
approximately  $0.3t_{\rm rh,i}$ (where $t_{\rm rh,i}$ is the 
initial half-mass relaxation time) to occur. The collapse time for other parameter choices is discussed at length in \cite{breenheggie1}.
In Fig. \ref{fig:BEHA} this occurs at $270$ N-body units. 
We will refer to this as the first collapse. This is followed by a phase of powered 
expansion. This can be seen in Fig. \ref{fig:BEHA} where, after the first 
collapse, both $r_{\rm c}$ and $r_{\rm h}$ increase up until $\sim 5000$ N-body units. As BH escape, 
the BH sub-system becomes less efficient at producing 
energy (see Section \ref{sec:limitations}) and the rate of expansion decreases. The core stops 
expanding and begins to contract again at $\approx 7500$ N-body units; at this stage there is only $15\%$ of the BH sub-system remaining (see Fig. \ref{fig:64kmassloss}, bottom, solid line). Most of the remaining BH escape before $\approx 9500$  N-body units leaving the system with a single remaining BH binary from $\sim 11500$ N-body units. The contraction of the core that begins after $\approx 7500$ N-body units shall be referred to as the second core collapse or recollapse. As with the first core collapse the core is contracting because there is not enough energy being produced to meet the energy demands of the cluster. As the core (which is dominated by the low mass stars  as most of the BH have escaped) becomes smaller towards the end of the run the remaining BH binary starts to interact strongly with the light stars, producing energy more efficiently. This causes the more rapid increase in $r_{\rm h}$ seen towards the end of the plot. The contraction of the core was still ongoing at the end of the run. The core will presumably continue to contract until balanced evolution is 
restored.
 If the last remaining BH binary is providing most of the energy, then how long that binary persists in the system depends on the hardness of that binary. Assuming the last BH binary is only slightly hard it is possible that the core contracts sufficiently for a single BH binary to produce the required energy to power the expansion of the system. Ultimately the BH binary will become hard enough to cause its ejection from the system. However if it is extremely hard it is likely that the binary gets ejected from the system during the second core collapse. If this happens the collapse of the core will continue until light binaries are produced as in a one-component model.

\subsection{The rate of loss of BH}\label{sec:nbody_M2dot}

\begin{table*}
\begin{center}
\caption{Results from N-body runs with parameters given in Table \ref{table:Ndetails}. The values given are stellar mass ratio $(m_2/m_1)$, initial total mass ratio $(M_2/M_1)$, initial total particle number $(N)$, initial number of BH $(N_2)$, core collapse time $(t_{\rm cc})$, the time at which $50\%$ ($T_{50\%}$) and $90\%$ ($T_{90\%}$) of the initial BH total mass has escaped, the recollapse time of the system, the rate of mass loss from the sub-system $-\dot{M}_2$, $\zeta$ (see Section \ref{sec:BHtwo}) and the number of the figure which plots the fraction of remaining BH mass ($M_2/M_{\rm 2,i}$) with time. Times for $T_{50\%}$, $T_{90\%}$ and the recollapse are given in N-body units and given in brackets in units of $t_{\rm rh,i}$. Times for $T_{50\%}$, $T_{90\%}$ and the recollapse are measured from the time at which core collapse finishes. The value of $T_{90\%}$ for 
the 
case $m_2/m_1=20$, 
$M_2/M_1 =0.01$ 
and $N=32k$ (marked with $*$) is actually the point where $88\%$ mass loss occurs; after this point all that remains in 
the system is a 
single binary BH. The values of $\dot{M}_2$ are given in units of $10^{-6}$ N-body units, (or $10^{-3}$ $t_{\rm rh,i}^{-1}$ for the values in brackets); these are measured between the loss of the first BH and  $50\%$ of the BH, by $\dot{M}_2 = -0.5M_{\rm 2,i}/T_{50\%}$. The values given in the subscript and superscript are the upper and lower $90\%$ confidence limits assuming that BH escape is a Poisson process. The values of $\zeta$ were measured by assuming $\dot{E}/|E| \approx \dot{r}_{\rm h}/r_{\rm h}$ (which holds if $|E| \propto GM^2/r_{\rm h}$ and $\dot{M}$ is small) and evaluating $\displaystyle{ \frac{0.138N}{\ln{\Lambda}} \frac{2}{3}\frac{d(r_{\rm h}^{\frac{3}{2}})}{dt} (\approx t_{\rm rh}\frac{\dot{r}_{\rm h}}{r_{\rm h}}}$  N-body units) between $2t_{\rm cc}$ and $t_{\rm cc}+T_{50\%}$. $\displaystyle{\frac{d(r_{\rm h}^{\frac{3}{2}})}{dt}}$ was evaluated by taking the slope of the best fit line to $r_{\rm h}^{\frac{3}{2}}$; the typical errors with the fitted lines were small, $<3\%$.} 
\begin{threeparttable}
\begin{tabular}{ c  |c |c|c|c|c|c|c|c|c|l|}
\bf{$\frac{m_{2}}{m_1}$}&$\frac{M_{2}}{M_1}$& $N$ & $N_2$ & $t_{\rm cc}$ &   $T_{50\%}$\tnote{a}            &    $T_{90\%}$\tnote{a}  & recollapse time\tnote{a}& $-\dot{M}_2$\tnote{b}  & $\zeta$ & Figures \\ \hline
$10$&$0.01$&$32k$ &$34$& $530$ &  $2070$ ($4.4$) &  $5610$ ($11.9$) & $8748$  ($18.6$) & $2.4^{3.6}_{1.5}$ ($1.1^{1.7}_{0.7}$)& $0.03$  & \ref{fig:32kM01massloss} (Top)\\ 
$20$&$0.01$&$32k$ &$17$& $380$ & $850$ ($1.8$) & $3000^*$ ($6.3^*$) & $> 8000$ ($17.0$) & $5.9^{10.6}_{2.9}$ ($2.8^{5.0}_{1.4}$) &  $0.06$ & \ref{fig:32kM01massloss} (Top)\\
%
$10$&$0.02$&$32k$&$66$& $492$  & $1658$ ($3.5$)& $5702$ ($12.1$) & $> 11876$ ($25.2$)  & $6.0^{8.1}_{4.4}$ ($2.8^{3.8}_{2.1}$) &$0.06$ &\ref{fig:32kM01massloss} (Bottom)\\ 
$20$&$0.02$&$32k$&$33$& $419$  & $940$  ($1.9$)&  $3816$ ($8.1$) &  $> 9902$ ($21.0$)  & $10.6^{16.0}_{6.8}$ ($5.0^{7.5}_{3.2}$) & $0.10$&\ref{fig:32kM01massloss} (Bottom)\\ 
%
%
%
$10$&$0.01$&$64k$  &$66$& $920$ & $3650$ ($4.2$)& $12950$ ($15.2$)    & $>13000$ ($15.3$)   & $1.4^{1.8}_{1.0}$  ($1.2^{1.6}_{0.9}$)& $0.03$&\ref{fig:64kmassloss} (Top)\\ 
$20$&$0.01$&$64k$  &$33$& $404$ & $1750$ ($2.1$)& $5400$ ($6.8$)      &  $> 7844$ ($9.2$)         & $2.9^{4.3}_{1.8}$ ($2.4^{3.6}_{1.5}$) & $0.05$&\ref{fig:64kmassloss} (Top) \\
%
$10$&$0.02$&$64k$  &$131$& $690$& $3230$ ($3.7$)& $12740$ ($14.6$)& $> 15580$ ($17.9$) & $3.5^{4.3}_{2.9}$  ($2.6^{3.2}_{2.1}$) & $0.05$&\ref{fig:64kmassloss} (Bottom)\\ 
$20$&$0.02$&$64k$  &$66$ & $270$& $2680$ ($3.2$)& $9495$ ($11.3$) & $> 19600$ ($23.4$) & $3.7^{5.0}_{2.7}$  ($3.2^{4.2}_{2.3}$) & $0.08$&\ref{fig:64kmassloss} (Bottom)\\ 
$20$&$0.02$&$128k$  &$131$ & $650$ & $4120$ ($2.7$) &  $>7076$ ($4.6$) & $>7076$ ($4.6$) & $2.4^{3.0}_{2.0}$  ($3.8^{4.6}_{3.0}$) &  $0.08$ & \ref{fig:128kmassloss}
\label{table:lifeN1}
\end{tabular}
     \begin{tablenotes}
       \item[a] Units: N-body units ($t_{\rm rh,i}$)
       \item[b] Units: $10^{-6}$ N-body units ($10^{-3}$ $t_{\rm rh,i}^{-1}$) 
     \end{tablenotes}
\end{threeparttable}
\end{center}
\end{table*}

\begin{figure}
\subfigure{\scalebox{0.60}{\includegraphics{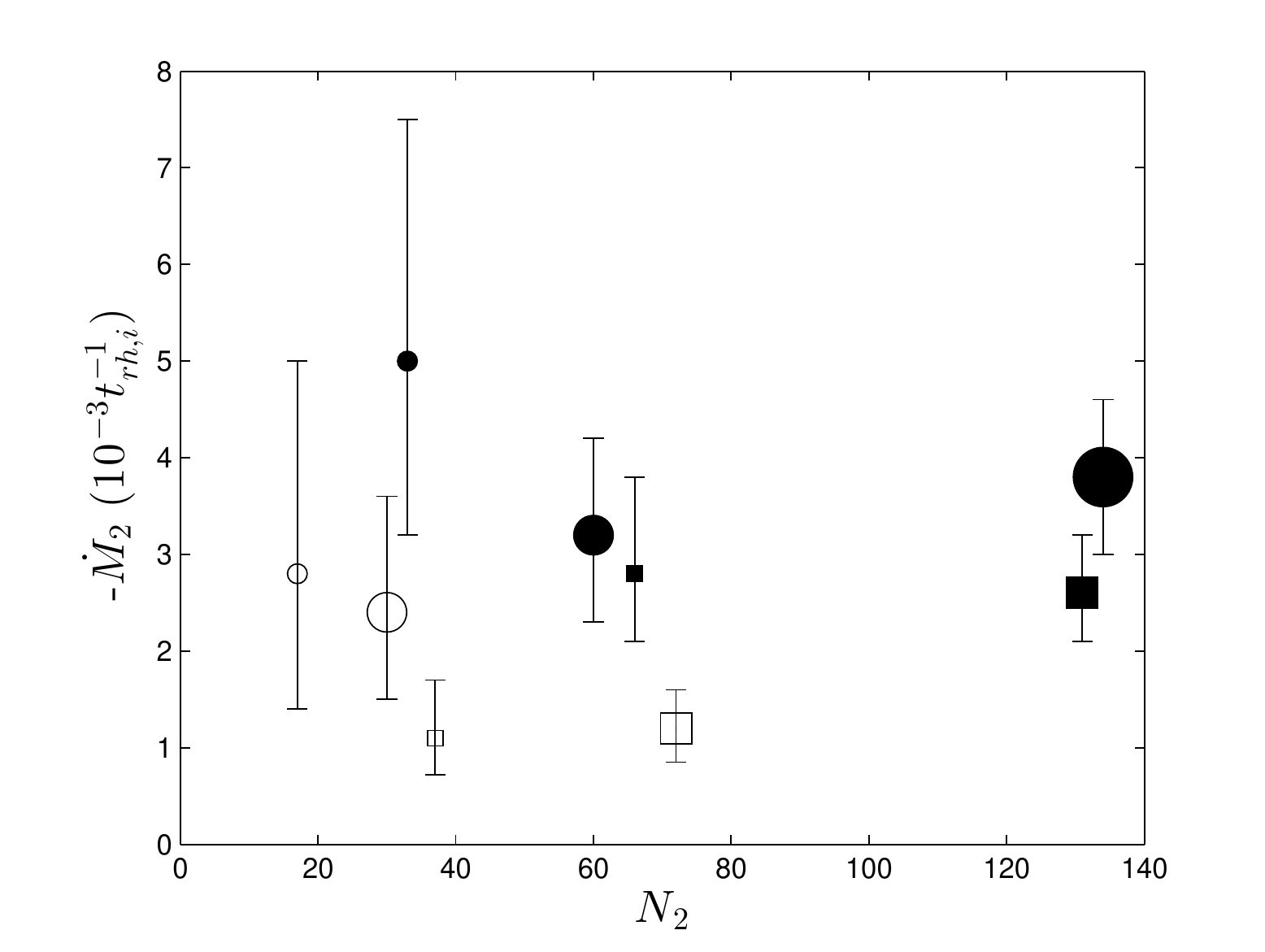}}}\quad
\caption{$\dot{M}_2$ (units of $10^{-3}t_{\rm rh,i}^{-1}$) versus initial value of $N_2$. Error bars indicate confidence limits (see Table 
\ref{table:lifeN1} for details). Circles represent $(20,M_2/M_1,N)$, squares represent $(10,M_2/M_1,N)$, filled symbols represent $(m_2/m_1,0.02,N)$, unfilled symbols represent $(m_2/m_1,0.01,N)$, larger symbols $(m_2/m_1,M_2/M_1,64k)$ (with the exception of the largest circle on the right which corresponds to $(20,0.02,128k)$) and smaller symbols $(m_2/m_1,M_2/M_1,32k)$. For cases with the same or similar initial values of $N_2$ some of the values were adjusted by $\lesssim 10\%$ to stop the symbols from overlapping. To a first approximation $\dot{M}_2$ is independent of $m_2/m_1$, $M_2/M_1$ and $N$.}
\label{fig:sctplot}
\end{figure}

Now we compare the values of $\dot{M}_2$ with the theory of Section \ref{sec:ej} (equation \ref{eq:M2dot}). $\dot{M}_2$ is estimated by 
calculating the average mass loss rate over the time taken (from the start of mass loss) for the BH sub-system to lose $50\%$ of its initial mass (i.e. $0.5M_{\rm 2,i}/T_{50\%}$;  where $T_{50\%}$ is the time taken from $t_{\rm cc}$ until $50\%$ of the BH sub-system has escaped, see Table \ref{table:lifeN1} for details). Note that there is a small systematic error introduced by measuring mass loss from the point at which it first occurs. The values of $\dot{M}_2$ (in units of $10^{-3}t_{\rm rh,i}^{-1}$)  are plotted in Fig. \ref{fig:sctplot}. The error bars (estimated as stated in the caption of Table \ref{table:lifeN1}) are large because $N_2$ is relatively small, and most data points are consistent with a value of approximately $3\times 10^{-3}t^{-1}_{\rm rh,i}$. Equation \ref{eq:M2dot}, with canonical values of $\alpha$, $\beta$ and $\zeta$ implies that the values in Fig. \ref{fig:sctplot} should be nearly $6.1 \times 10^{-3}t^{-1}_{\rm rh,i}$ and we will now consider reasons for the discrepancy.


\begin{figure}
\subfigure{\scalebox{0.60}{\includegraphics{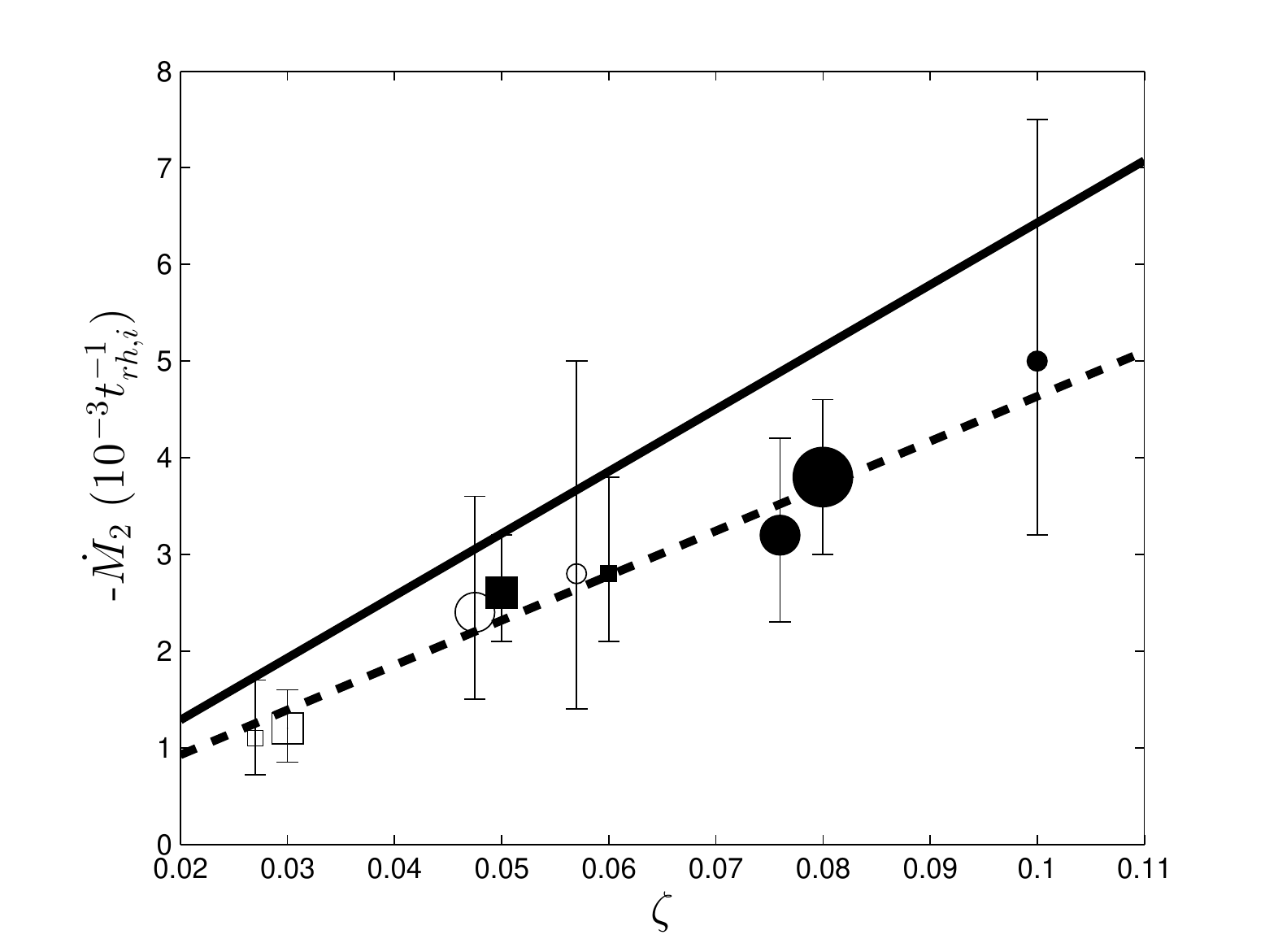}}}\quad
\caption{$\dot{M}_2$ (units of $10^{-3}t_{\rm rh,i}^{-1}$) versus value of $\zeta$; error bars indicate $90\%$ confidence limits (see Table \ref{table:lifeN1} for details). The symbols represent the same runs as in Fig. \ref{fig:sctplot}, the solid line represents the predicted values of $\dot{M}_2$ using the value of $\alpha/\beta = 0.068$ (based on theoretical arguments), and the dashed line represents the predicted values of $\dot{M}_2$ using the value of $\alpha/\beta = 0.051$ (the empirical value). For cases with the same or similar values of $\zeta$ some of the values of $\zeta$ were adjusted by $ \pm 5\%$ to stop the symbols from overlapping. See text for details.}
\label{fig:sctplotb}
\end{figure}

First, this estimate can be improved upon by taking into account the fact that the system expands as energy is being generated, increasing the relaxation time and in turn decreasing the mass loss rate. The improved estimate can be calculated by using equation \ref{eq:masslosslog} to estimate the time taken for half the BH to be lost and evaluating $\dot{M}_2$ as was done in Table \ref{table:lifeN1}. This results in a slightly smaller estimate of $5.4\times 10^{-3}t^{-1}_{\rm rh,i}$, which is still significantly larger than most of the values in Table \ref{table:lifeN1}.

Another factor is that the values of $\zeta$ for most of the runs are smaller than the canonical value used for the estimate (i.e. $\zeta \approx 0.09$). Equation \ref{eq:masslosslog} predicts an approximately linear dependence of $\dot{M}_2$ on $\zeta$, and this is clearly confirmed in Fig. \ref{fig:sctplotb}. The solid line, which represents the predicted values of $\dot{M}_2$ with varying $\zeta$, nevertheless lies above all the numerical results and outside the confidence intervals for all but a few of the runs. However by adjusting the value of $\alpha/\beta$ (in equations \ref{eq:M2dot} and \ref{eq:masslosslog} $\alpha$ and $\beta$ only appear in the form $\alpha/\beta$) from  $\alpha/\beta \approx 0.068$ (the value estimated on the basis of theoretical arguments) to  $\alpha/\beta \approx 0.051$ the theory comes into very good agreement with the values of $\dot{M}_2$. This can be seen in Fig. \ref{fig:sctplotb} where the dashed line represents the predicted values of 
$\dot{M}_2$  based on a value of  $\alpha/\beta \approx 0.051$. The discussion of Section \ref{sec:ej} makes it clear that the canonical values of $\alpha$ and, especially, $\beta$ are subject to uncertainty, the latter resting entirely on approximate theoretical arguments. The suggested revision of $\alpha/\beta$ 
cannot be ruled out on these grounds.  

\begin{figure}
\subfigure{\scalebox{0.75}{\includegraphics{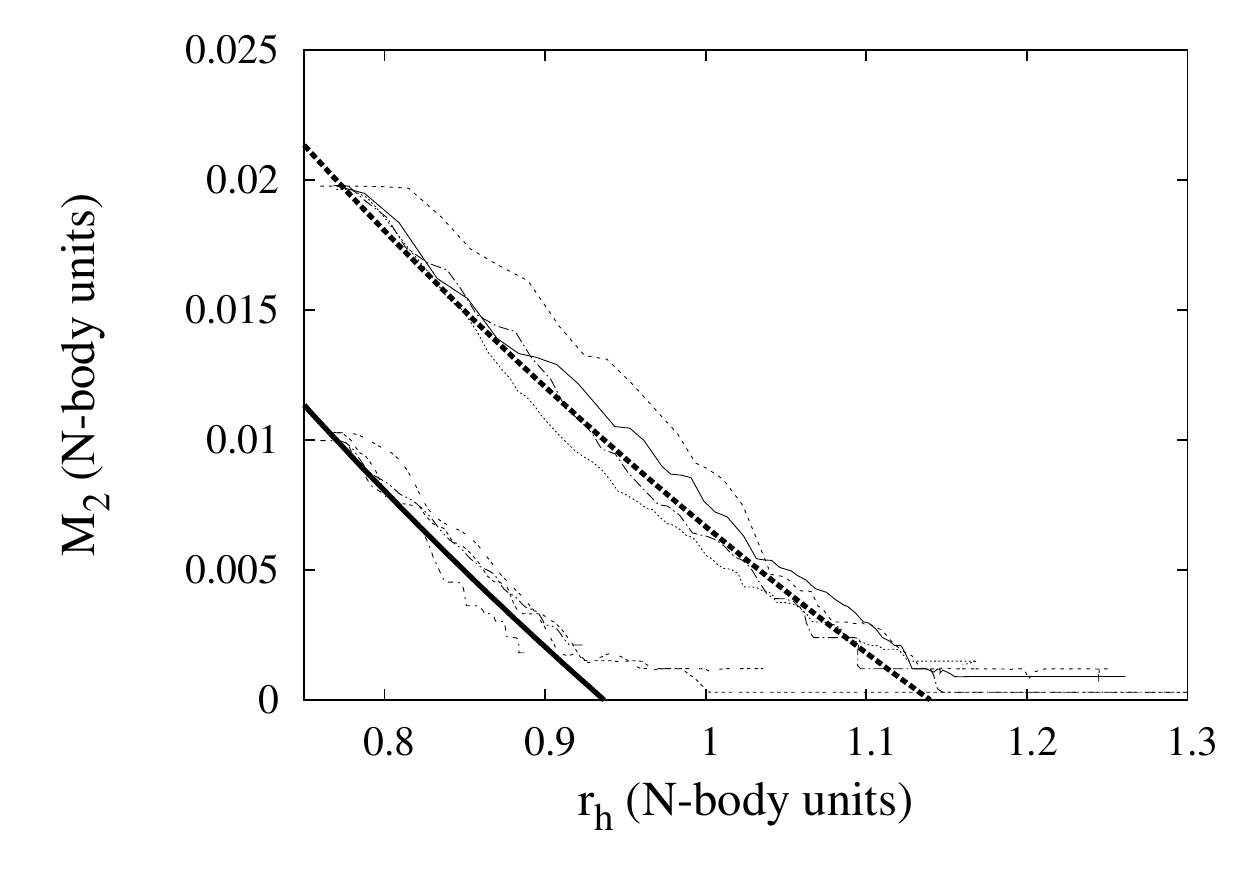}}}
\caption{Evolution of $M_2$ vs $r_{\rm h}$ for the $N=32k$ and $64k$ models in Table \ref{table:lifeN1}. The thick dashed line is the theoretical prediction (see equation \ref{eq:MvsRH}) for the initial value of $M_2/M_1=0.02$ and the thick solid line is the theoretical prediction for the initial value of $M_2/M_1=0.01$. The value of $r_{\rm h,i}$ used in equation \ref{eq:MvsRH} was $0.77$ and the empirical value of  $\alpha/\beta \approx 0.051$ was used for all models. In all cases the behaviour of the N-body runs is in approximate quantitative agreement with the predicted behaviour until there are only a few BH remaining.}
\label{fig:BHMRrelate}
\end{figure}

Equation \ref{eq:MvsRH} allows us to test the theory constructed in Section \ref{sec:ej}, in a $\zeta$-independent way. In Fig. \ref{fig:BHMRrelate} the observed dependence of $M_2$ on $r_{\rm h}$ is in satisfactory agreement with the predictions based on equation \ref{eq:MvsRH}.

\begin{figure}
\subfigure{\scalebox{0.6}{\includegraphics{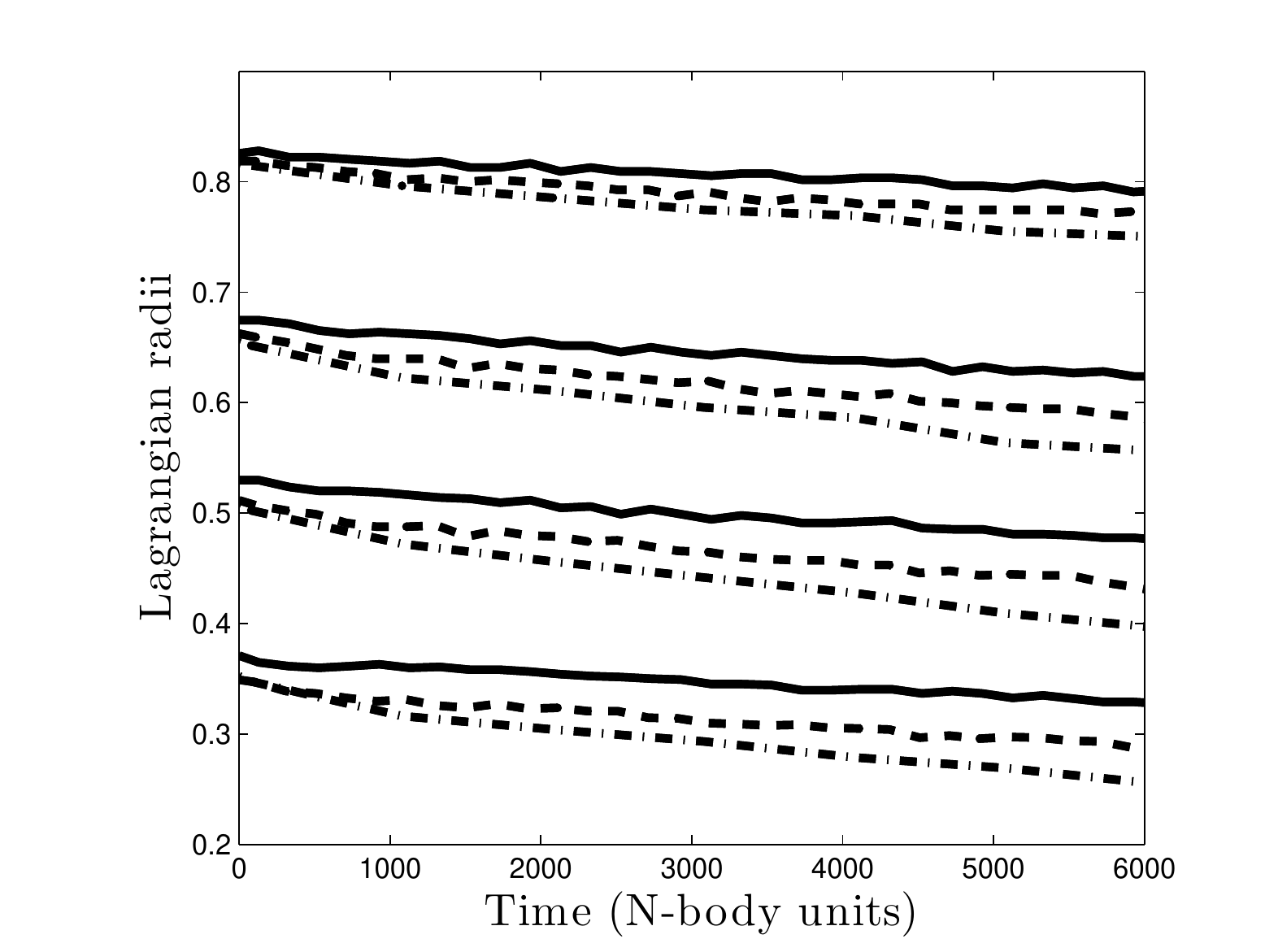}}}
\caption{Relative contraction in Lagrangian radii ($40\%$, $30\%$, $20\%$ and $10\%$) over time (N-body units) for ($20$,$0.02$,$64k$) solid line, ($10$,$0.02$,$64k$) dashed line  and  ($10$,$0.01$,$64k$) dot dash line. $T=0$ in all models is set at the time mass loss starts from the BH sub-system. Radii are measured in units of the half-mass radius. The values of $\zeta$ for the three runs are $0.08$, $0.05$ and $0.03$, respectively. The parameter $\zeta$ measures the dimensionless expansion rate of the half-mass radius and the figure shows that slow expansion is  associated with relative contraction of the inner Lagrangian radii.}
\label{fig:BHkh}
\end{figure}

The lower values of $\zeta$ in Table \ref{table:lifeN1} may result from systems in which the BH sub-system is incapable of producing the required energy for the system to achieve balanced evolution. This could possibly be due to the small values of $N_2$; most of these models reach values of $N_2$ (at time $T_{50\%}$) below the point at which the theory of Section \ref{sec:ej} is expected to apply (see Section \ref{sec:limitations}). If a system is not in balanced evolution it is expected to undergo contraction of the inner Lagrangian radii relative to $r_{\rm h}$, qualitatively as in conventional core collapse. This is illustrated in Fig. \ref{fig:BHkh} for three of the N-body runs in Table \ref{table:lifeN1}. The systems with smaller values of $\zeta$ show greater contraction. Note that the values $\zeta$ are only evaluated over the period to $T_{50\%}$ and appear to decrease after $T_{50\%}$ ($\sim 3000$). Indeed this is what would be expected as the expansion is affected by the weakening energy generation (see 
also Fig. \ref{fig:BEHA}).


In this section we have assumed that mass segregation concentrates the BH in the centre of the system by the time of the first core collapse. Though this prevents further contraction of the central BH sub-system, this is not true of all the BH, as discussed in Appendix \ref{sec:Heatingouterlagra} and \cite{Morscher2012}. The outermost BH can continue contracting after core collapse has occurred, indicating that mass segregation can continue in the outermost parts of a system for a while after core collapse. This results in additional heating which is not associated with energy production. The effect of this heating is expected to be small for the models in Table \ref{table:Ndetails} due to the small particle number, although the effect may be more significant in larger systems and may be enhanced by the presence of a mass spectrum.

\subsection{Lifetime of BH sub-systems}
Now that the dependence of $\dot{M}_2$ on cluster properties has been discussed, it is natural to move on to considering how long the  BH sub-system lasts. For this purpose we shall define the life time of the BH sub-system as the time taken from the core collapse of the  BH sub-system  (which occurs at $t_{\rm cc}$) until the BH sub-system has lost $90\%$ of its initial mass ($T_{90\%}$). These values are given in Table \ref{table:lifeN1}, where it can be seen that $T_{90\%} \sim 10t_{\rm rh,i}$. Equation \ref{eq:masslosslog} in Section \ref{sec:ej} can be used to estimate the life time of the sub-system (using $\alpha/\beta =0.051$). For $M_2/M_1=0.01$ the theory predicts $T_{90\%} \sim 2.2 t_{\rm rh,i}$ and $T_{90\%} \sim 5.2t_{\rm rh,i}$ for $M_2/M_1=0.02$. These values are significantly smaller than the values seen in Table \ref{table:lifeN1}. As stated in the previous subsection most of the values of $\zeta$ in Table \ref{table:lifeN1} are below the value used in Section \ref{sec:ej}. 
Adjusting $\zeta$ to $0.05$ in equation \ref{eq:masslosslog} increases the predicted values of $T_{90\%}$ to $4.0t_{\rm rh,i}$ for $M_2/M_1=0.01$ and $9.3t_{\rm rh,i}$ for $M_2/M_1=0.02$. These values are still significantly smaller than those given in Table \ref{table:lifeN1} with the corresponding value of $\zeta$. The difference between the empirically found values and the theoretical estimates might be accounted for by the fact that the theory assumes a constant value of $\zeta$; however $\zeta$ is expected to decrease as the BH sub-system evaporates (see Section \ref{sec:nbody_intro}). This behaviour is illustrated by the behaviour of $r_{\rm h}$ in Fig. \ref{fig:BEHA}. There is a hint that the same decrease of $\zeta$ with decreasing $N$ may also be present in one-component models \citep{AlexanderGieles2012}. Also as can be seen from the values of $\zeta$ in Table \ref{table:lifeN1} some systems appear to be incapable of achieving balanced evolution at any time throughout the loss 
of the BH sub-system. 

The expected evolution of a BH sub-system can be illustrated using Fig. \ref{fig:sctplotb}. If we assume that the BH sub-system is capable of achieving balanced evolution, after the formation of the BH sub-system $\zeta$ and $-\dot{M}_2$ are predicted to rapidly reach the balanced evolution values of $\zeta\approx 0.09$ and $-\dot{M}_2\approx 4 \times 10^{-3}t^{-1}_{\rm rh,i}$ (just to upper right of the large filled circle). As the BH sub-system loses mass it will eventually reach the point where it is no longer capable of generating the energy needed for balanced evolution. After this the system will move down the dashed line towards the origin. As it does so the rate of mass loss decreases, prolonging the life of the BH sub-system. This picture may explain the longer lifetimes given in Table \ref{table:lifeN1} and is consistent with the evolution of $r_{\rm h}$ in Fig. \ref{fig:BEHA}.

Finally we briefly consider the recollapse time of these systems (see Table \ref{table:lifeN1}). This is the time between the first and second core collapse. It can be interpreted as approximately the time it takes for the system to achieve balanced evolution once the BH sub-system has been exhausted. Most of the N-body runs do not reach the second core collapse, and therefore mostly lower limits on the recollapse time are given in Table \ref{table:lifeN1}. From the results in Table \ref{table:lifeN1} this time is at least roughly the same time as the core collapse time of a one component Plummer model 
\citep[$\approx 15t_{\rm rh,i}$ see][]{HeggieHut2003} but can be longer because of the offsetting effect of BH heating.

\section{Gravothermal Oscillations}\label{sec:GTO}

The conditions for the onset of gravothermal oscillations in two-component models have been studied by \cite{breenheggie1}, who found that the value of $N_2$ (the number of heavy stars) could be used as an approximate stability condition (where the stability boundary is at $N_2\sim3000$) for a wide range of stellar and total mass ratios ($2\le m_2/m_1 \le 50 $ and $0.1 \le M_2/M_1 \le 1.0$). \cite{breenheggie2}, who researched the onset of gravothermal oscillation in multi-component systems, found that the parameter called the effective particle number $N_{\rm ef}$ (defined as $M/m_{\rm max}$) could be used as an approximate stability condition for both the multi-component systems they studied and the two-component models of \cite{breenheggie1}. The stability boundary they found was at $N_{\rm ef} \sim 10^4$, which is also consistent with the stability boundary of the one-component model at $N=7000$ \citep{Goodman1987}. However both those stability conditions relied on the assumption that the 
heavy 
component (or heavier stars for the multi-component case) dominated the evolution of the system, in the sense that 
 the heavy component determined the rate of energy generation. This is not the case for the systems considered in the present paper as the total mass in the heavy component is so small,  and so $N_{\rm ef}$ will not be considered further. (In the systems considered in this paper the heavy component dominates the production of energy, but the light component controls how much energy is created.) We will now investigate the onset of gravothermal oscillation for the systems of interest in the present paper (where $M_2/M_1 \ll 1.0$ and $m_2/m_1 \gtrsim 5$).

The critical number of stars $(N_{\rm crit})$ at which gravothermal oscillations first manifest was found for the gas models with $M_2/M_1=0.05,0.01$ and, $m_2/m_1 = 5,$ $10,$ $20,$ and $50$ (see Section \ref{sec:rhsec}). The results are given in Table \ref{table:GTO}. The values of $N_2$ at $N_{\rm crit}$ for the runs in Table \ref{table:GTO} are given in Table \ref{table:GTO2}. The values for $M_{2}/M_1 =0.5$ and $0.1$ from \cite{breenheggie1} have also been  included in Tables \ref{table:GTO} and \ref{table:GTO2} for reference. For fixed $m_2/m_1$, the system becomes unstable at a roughly fixed value of $N_2$ ($N_2\approx 1500$ for $M_2/M_1=0.05$ and $N_2\approx900$ for $M_2/M_1=0.01$). These results suggest that $N_2$ still provides an approximate stability condition (for fixed $M_2/M_1$) even for models in which the heavy component only makes up a tiny fraction of the system. 

\begin{figure}
\subfigure{\scalebox{0.60}{\includegraphics{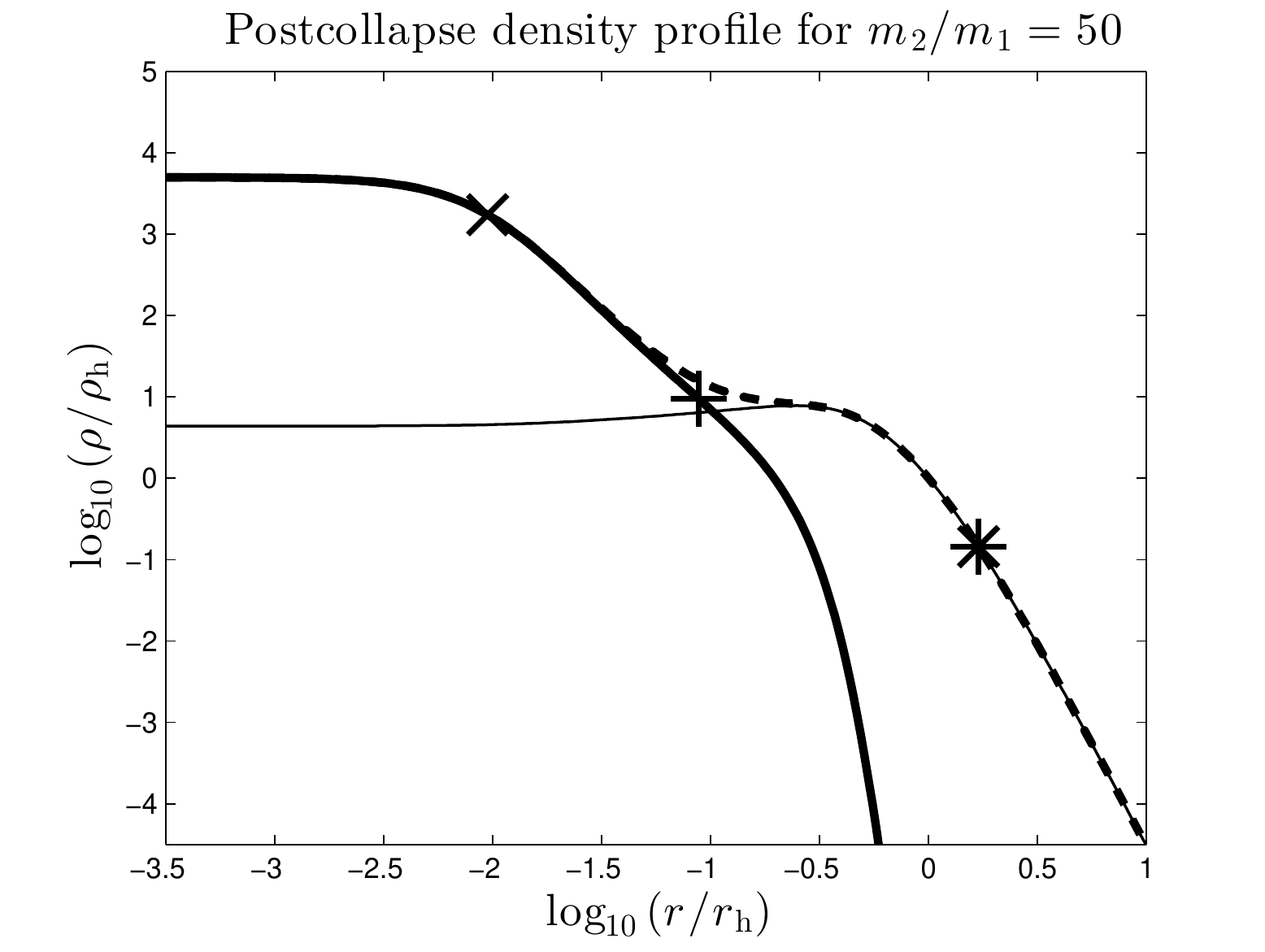}}}\quad
\subfigure{\scalebox{0.60}{\includegraphics{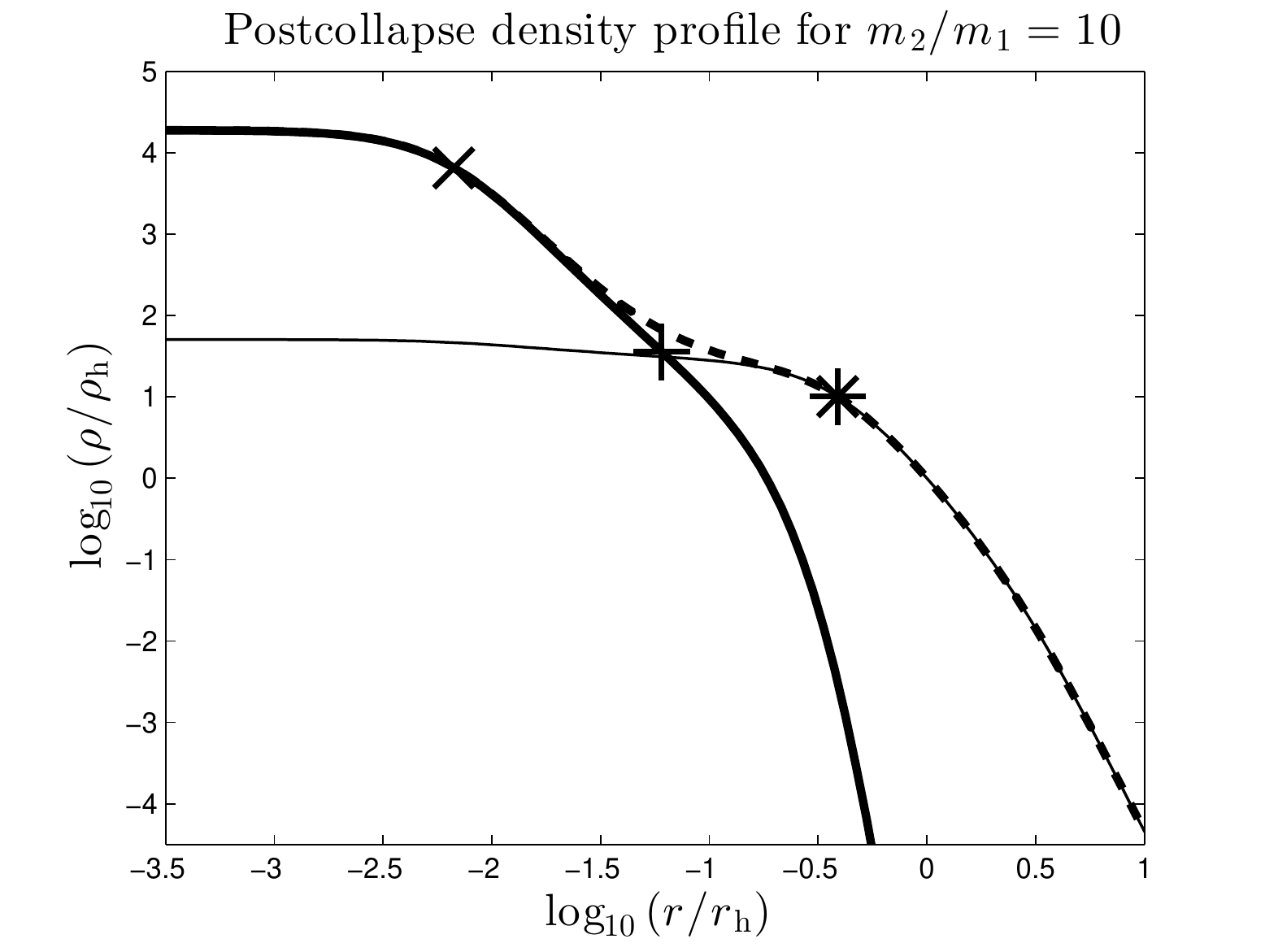}}}
\caption{Postcollapse density profile in gas models of two-component systems with $M_2/M_1=0.01$ near the onset of gravothermal oscillations for $m_2/m_1=50$ ($N=4.3\times10^6$, top) and $m_2/m_1=10$ ($N=8.5 \times 10^5$, bottom). The following is shown in the plot: $\rho_{1}$ (thin line), $\rho_{2}$ (thick line), $\rho_{\rm tot}$ (dashed line), core radius of heavy component $r_{\rm c,2}$ ($\times$), core radius of light component $r_{\rm c,1}$ ($\ast$) and $r_{\rm h,2}$ ($+$). The core radii have been defined as $r_{{\rm c},i}=\sqrt{9\sigma_{{\rm c},i}^2/(4\pi\rho_{{\rm c},i})}$. For the case of $m_2/m_1=50$, the BH sub-system creates a density hole in the light component: the density of lights in the centre is approximately a factor of $2$ less then its highest value (which occurs at $\log{r/r_{\rm h}} \simeq -0.5$). The remarkably large value of $r_{\rm c,1}$ for this case results from the low value of $\rho_{\rm c,1}$ and a high value of $\sigma_{\rm c,1}^2$ caused by the presence of the BH sub-system.}
\label{fig:densitypro}
\end{figure}

Given that the theory in the present paper is built around the assumption that the light component determines the evolution of the BH sub-system, it may be surprising that the appearance of gravothermal oscillations seems solely determined (for fixed $M_2/M_1$) by the number of stars in the heavy component. However this may be explained if one considers the unique structure of systems containing a BH sub-system. The presence of a BH sub-system tends to produce  a system with two cores, one small BH core and another much larger light core, which is larger than the half mass radius of the BH sub-system (\citealt{Merritt2004}; \citealt{Mackey2007}; also see Fig. \ref{fig:densitypro} in the present paper). As gravothermal oscillation in a one-component system requires a small core to half mass ratio \citep{Goodman1987}, the light system itself is expected to be highly stable against gravothermal oscillations. However as the BH sub-system has to meet the energy generation requirements of the entire system, 
the BH 
sub-system can have a very small ratio of core radius to half mass radius (see Section \ref{sec:BHtwo}). If the onset of gravothermal oscillation is a result of the  BH sub-system itself becoming unstable then it would be expected that the BH sub-systems would have similar structure at the stability boundary. This is indeed the case as can be seen in Fig. \ref{fig:densitypro} which shows the post collapse density profile of two  systems (with $m_2/m_1=50$ and $m_2/m_1=10$) near the stability boundary (i.e. $N$ is slightly  smaller than $N_{\rm crit}$): the profiles of the heavy component are almost identical (in terms of density contrast, i.e. $\rho_{\rm 2}/\rho_{\rm h,2}$) whereas the profiles of the light component are significantly different. In fact if one were to plot the BH sub-systems in units of $\rho_{\rm c,2}$ vs $r_{\rm c,2}$ the BH sub-systems would be nearly indistinguishable. We study this more quantitatively below.

\begin{table}
\begin{center}
\caption{Critical values of $N$ in units of $10^4$ }
\begin{tabular}{ c  |c |c| c |c |}
 \bf{$\frac{m_{2}}{m_1}\backslash \frac{M_2}{M_1}$ }    &  \bf{0.5}   & \bf{0.1}   & \bf{0.05}   & \bf{0.01}     \\ \cline{1-5} 
 \bf{50}         &  30  & 100   &  130     &  450 \\ 
 \bf{20}         &  13  &  36   &  63      &  180 \\
 \bf{10}         &  7.2 &  22   &  32      &  90  \\ 
 \bf{5}          &  4.0 &  10   &  17      &  40   \\ 
%
\label{table:GTO}
\end{tabular}
\end{center}
\end{table}

\begin{table}
\begin{center}
\caption{Values of $N_2$ at $N_{\rm crit}$ in units of $10^3$ }\label{table:GTO2}
\begin{tabular}{ c  |c |c| c |c |}
 \bf{$\frac{m_{2}}{m_1}\backslash \frac{M_2}{M_1}$ }    &  \bf{0.5}   & \bf{0.1}   & \bf{0.05}   & \bf{0.01}     \\ \cline{1-5} 
 \bf{50}         &  3.0 & 2.0   &  1.3     &  0.9 \\ 
 \bf{20}         &  3.2 & 1.8   &  1.6     &  0.9 \\
 \bf{10}         &  3.4 & 2.2   &  1.6     &  0.9 \\ 
 \bf{5}          &  3.6 & 2.0   &  1.7     &  0.8   \\ 
%
\end{tabular}
\end{center}
\end{table}

The Goodman stability parameter \citep{Goodman1993} (or a somewhat modified version \citep{breenheggie1,breenheggie2}) has been found to provide a stability criterion. The Goodman stability parameter is defined as $$\epsilon \equiv \frac{E_{\rm tot}/t_{\rm rh} }{E_{\rm c} /t_{\rm rc}},$$ where $E_{\rm c}$ is the energy of the core. The critical value for the one-component model is $\log_{10} \epsilon \approx -2$. This condition was also found to apply for the Spitzer stable two-component models studied by \cite{KimLeeGood1998}. However, \cite{breenheggie1} found the critical value of $\epsilon$ to vary for the Spitzer unstable models they studied. They found that by slightly modifying the definition  of $\epsilon$ ($\epsilon_2$) a much improved stability criterion could be found, with a critical value $\log_{10} \epsilon_2 \approx -1.5$. We can test a version of these parameters for the BH sub-systems by suitably modifying $\epsilon$ to
$$\epsilon_{\rm BH} \equiv \frac{E_{\rm BH}/t_{\rm rh,2} }{E_{\rm c,2} /t_{\rm rc,2}}$$
where $E_{\rm c,2}$ and $E_{\rm BH}$ are the energy of the BH core and the total energy of the BH sub-system respectively. $\epsilon_{\rm BH}$ was measured for a range of systems with $M_2/M_1=0.01$ and the results are presented in Table \ref{table:rh2rcEPS} along with the values of $r_{\rm c,2}/r_{\rm h,2}$ (following the discussion of the previous paragraph). As can be seen in Table \ref{table:rh2rcEPS} all the systems with large enough $m_2/m_1$ have similar values of $\log_{10}\epsilon_{\rm BH}$ and $r_{\rm c,2}/r_{\rm h,2}$ at their corresponding value of $N_{\rm crit}$ which supports the assertion that the onset of gravothermal oscillation depends on the structure of the BH sub-system. 

The critical values of  $\log_{10}\epsilon_{\rm BH}$ are larger than the values of $\log_{10}\epsilon$ found for two-component models by \cite{KimLeeGood1998} and \cite{breenheggie1}, and that found for the one-component model by \cite{Goodman1993} by approximately 1.7 dex. Also the critical values of $r_{\rm c,2}/r_{\rm h,2}$  are larger than the corresponding critical value for a single-component system \citep{Goodman1987}, i.e. $r_{\rm c}/r_{\rm h} \approx0.02$. This may be because the maximum radius of the isothermal region in the BH sub-system is larger than $r_{\rm h,2}$, as was hinted by \cite{breenheggie1}. If the condition for gravothermal instability is that the density contrast across the isothermal region exceeds some critical value, and if the edge of this region is well outside $r_{\rm h,2}$, then it can be understood why the critical value of $r_{\rm c,2}/r_{\rm h,2}$ is larger than Goodman's value. To investigate this, the size of the isothermal region was measured for ($50$,$0.01$,$4.3\times10^6$),
which is shown in Fig. \ref{fig:densitypro} (Top). The edge of the isothermal region ($r_{\rm iso}$) was defined as the radius at which $\sigma_2^2$ reaches $80\%$ of it central value.  This gives  $r_{\rm c,2}/r_{\rm iso}\approx 0.022$ which is consistent with the value of $r_{\rm c}/r_{\rm h}$ found by \cite{Goodman1987} for  the one-component model. (For a one-component gas model, $r_{\rm c}/r_{\rm iso}\simeq 0.016$ near the stability boundary.)


\begin{table}
\begin{center}
\caption{Values of $\log_{10}\epsilon_{\rm BH}$ and $r_{\rm c,2}/r_{\rm h,2}$ near $N_{\rm{crit}}$ for systems with $M_2/M_1=0.01$. For the corresponding values of $N_{\rm crit}$ see Table \ref{table:GTO}. The results in this table indicate that gravothermal oscillation manifests once a certain value of $r_{\rm c,2}/r_{\rm h,2}$ (or  $\log_{10} \epsilon$) is reached. See text for details.}\label{table:rh2rcEPS}
\begin{tabular}{ccccc}
$m_2/m_1$                   & $5$      & $10$     & $20$      &  $50$  \\
$r_{\rm c,2}/r_{\rm h,2}$           &  $0.14$  & $0.12$   & $0.12$    & $0.11$   \\ 
$\log_{10} \epsilon_{\rm BH}$        &  $-0.21$  &  $-0.30$&   $-0.31$    & $-0.32$  
\end{tabular}
\end{center}
\end{table}

\section{Conclusion and Discussion}\label{sec:conanddis}
\subsection{Summary} 
In this paper we have studied systems intended to resemble those containing a significant population of black holes (BH), i.e. two-component systems with one component being the BH and the other the rest of the stars in the system. It was argued in Section \ref{sec:evap} that mass loss by evaporation due to two body relaxation in the BH sub-system does not cause significant mass loss of BH and can be neglected. The  principal  mechanism for removing BH, for the models considered in the present paper, is superelastic encounters involving BH binaries and single BH in the core of the BH sub-system. By considering these systems to be in balanced evolution, predictions were made regarding the BH sub-system, for example the escape rate of BH (see Section \ref{sec:ej}). Some of the potential limitations of the theory were also discussed in Section \ref{sec:limitations}. 

The theory in Sections \ref{sec:theory} and  \ref{sec:bal} makes predictions about the structure of the BH sub-system, particular regarding the variation of  $r_{\rm c,2}/r_{\rm h,2}$ with $m_2/m_1$ and $M_2/M_1$ (see equation \ref{eq:balone}). (Here the subscripts 2 and 1 refer to the BH and the other stars, $M$, $m$ denoted the total and individual masses, and $r_{\rm c}$, $r_{\rm h}$ the core and half-mass radii, respectively). The theory was tested in Section \ref{sec:rhsecgas} with gas models and was found to be in good agreement with theory under the condition that $m_2/m_1 \gtrsim 10$ and that $N \gtrsim 128k$  (see Figs \ref{fig:rBHrh2vari} and \ref{fig:rBHrh2variM}).  The disagreement with the theory outside of those conditions may be attributable to small $N_2$ and the fact that the systems become Spitzer stable at low $m_2/m_1$. One of the assumptions of the theory was that the BH sub-system only made a small contribution to the central potential (see Section \ref{sec:ej}), which was tested in Section 
\ref{sec:phi} by measuring the contribution to the central potential of each component in a series of simulations. 

It was argued in Section \ref{sec:ej}  that the rate of mass loss from the $BH$ sub-system should be approximately independent of the properties of the BH sub-system (i.e. $M_2/M_1$ and $m_2/m_1$). This theory only requires that the light component regulates the rate of energy production and does not rely on the stronger assumption that energy is transported through the BH sub-system as outlined in Section \ref{sec:theory}. In Section \ref{sec:nbody_M2dot} the results of a number of N-body runs were presented (see Table \ref{table:Ndetails} and Table \ref{table:lifeN1}) and the results were used to test the predicted mass loss rates from Section \ref{sec:ej}. With the exception of systems with $m_2/m_1=10$ and $M_2/M_1=0.01$, the mass loss rates for the BH sub-system were all consistent with a value of $\dot{M}_2 \sim 3.0\times 10^{-3}t_{\rm rh}^{-1}$, where $t_{\rm rh}$ is the half-mass relaxation time of the entire system. However most of the runs had a lower value of the dimensionless expansion rate $\zeta$ than 
expected and $\dot{M}_2$ was found to vary approximately linearly with $\zeta$ (see Fig. 
\ref{fig:sctplotb}). Once the variation of $\zeta$ was taken into account and the value of $\alpha/\beta$ (where $\alpha$, $\beta$ are dimensionless parameters determining the energy and central potential) was adjusted to $0.051$ (see Sections \ref{sec:ej} and \ref{sec:nbody_M2dot}), there was good agreement between the empirical values of $\dot{M}_2$ in Table \ref{table:lifeN1} and the predicted values of $\dot{M}_2$ made using equation \ref{eq:masslosslog}. The low values of $\zeta$ seen in Table \ref{table:lifeN1} may result from the small number of BH in these systems which possibly results in the inability of the BH sub-system to maintain balanced evolution. Larger simulations will be required before this explanation can be confirmed.

In Section \ref{sec:GTO} we considered gravothermal oscillations in systems containing a BH sub-system. This extends the parameter space of two-component clusters studied by \cite{breenheggie1} to lower values of $M_2/M_1$ for $m_2/m_1\ge5$. The results in this section imply that the gravothermal instability manifests when the BH sub-system reaches a certain profile (see Fig. \ref{fig:densitypro} and Table \ref{table:rh2rcEPS}). A version of the Goodman stability parameter was also tested for systems with  $M_2/M_1=0.01$ and was found to provide an  approximate stability condition, although the critical  value was  significantly larger than the value measured for one-component models. The difference between the critical values for the BH sub-systems and the one-component models may result from the fact that the BH sub-system is approximately isothermal to larger radii than $r_{\rm h,2}$. The ratio between $r_{\rm c,2}$ and the  radius of the isothermal region ($r_{\rm iso}$)  was measured for a selected model 
and 
was 
found to be consistent with the critical value of 
$r_{\rm c}/r_{\rm iso}$ found for a one-component system. These results indicate that the onset of gravothermal oscillation for systems containing a BH sub-system is determined by the properties of the BH sub-system.

\subsection{Astrophyical issues}
In the present paper we have made several simplifying assumptions. Importantly we have ignored stellar evolution and the effect of a mass spectrum. A mass spectrum can increase the rate of evolution of a system \citep{Gieles_et_al2010}, which by the theory in Section \ref{sec:ej} would lead to a faster escape rate of BH. On the other hand mass loss via stellar evolution from the formation of the BH can cause the system to expand \citep{Mackey2007} increasing the relaxation time in the system. This in turn would reduce the rate of energy generation in the system, which by the theory in Section \ref{sec:ej} would prolong the life of the BH sub-system. 

Another simplifying assumption was not to consider the removal of BH by natal kicks, which if large could significantly reduce the retained BH population. The topic of natal kicks for black holes is still under debate, so here we will only give the topic very general consideration. The ejection of BH by natal kicks is itself an energy source, which heats the system in qualitatively the same way as a BH ejected by  superelectic encounters with binaries. Also if a natal kick is not significant enough to remove a BH from the system the BH would shed much of the kinetic energy gained from the kick to the other stars in the system. This is analogous to the results of \cite{Fregeau2}, who found that adding natal kicks to white dwarfs was an  additional energy source.  

Another topic we have ignored is the presence of more than one stellar population in many globular clusters. In the typical scenario for the formation of the second generation stars in a globular cluster \citep{Ventura} ejecta from asymptotic giant branch stars cools and collects in the centre of the cluster. If a BH sub-system is  already  present at the centre this could lead to a significant increase in the stellar mass of the BH \citep{Krause2012,Leighetal2012} and by the theory in Section \ref{sec:BHtwo} an increase in the total mass of the BH sub-system would increase its life time. Even in a single population scenario physical collisions can occur between BH and other stars in the system \citep{Gierszetal}, and this also would increase the total mass in the BH sub-system.

We have not considered systems which contain an intermediate mass black hole (IMBH) alongside a BH sub-system. A recent radio survey by \cite{Strader2012} found no evidence of IMBH in the three globular clusters M15, M19, and M22. However, there could be other clusters which contain both an IMBH and a BH sub-system and this would be an interesting topic for future work.

\subsection{Classification of two-component systems}\label{sec:classification}

\begin{figure}
\subfigure{\scalebox{0.60}{\includegraphics{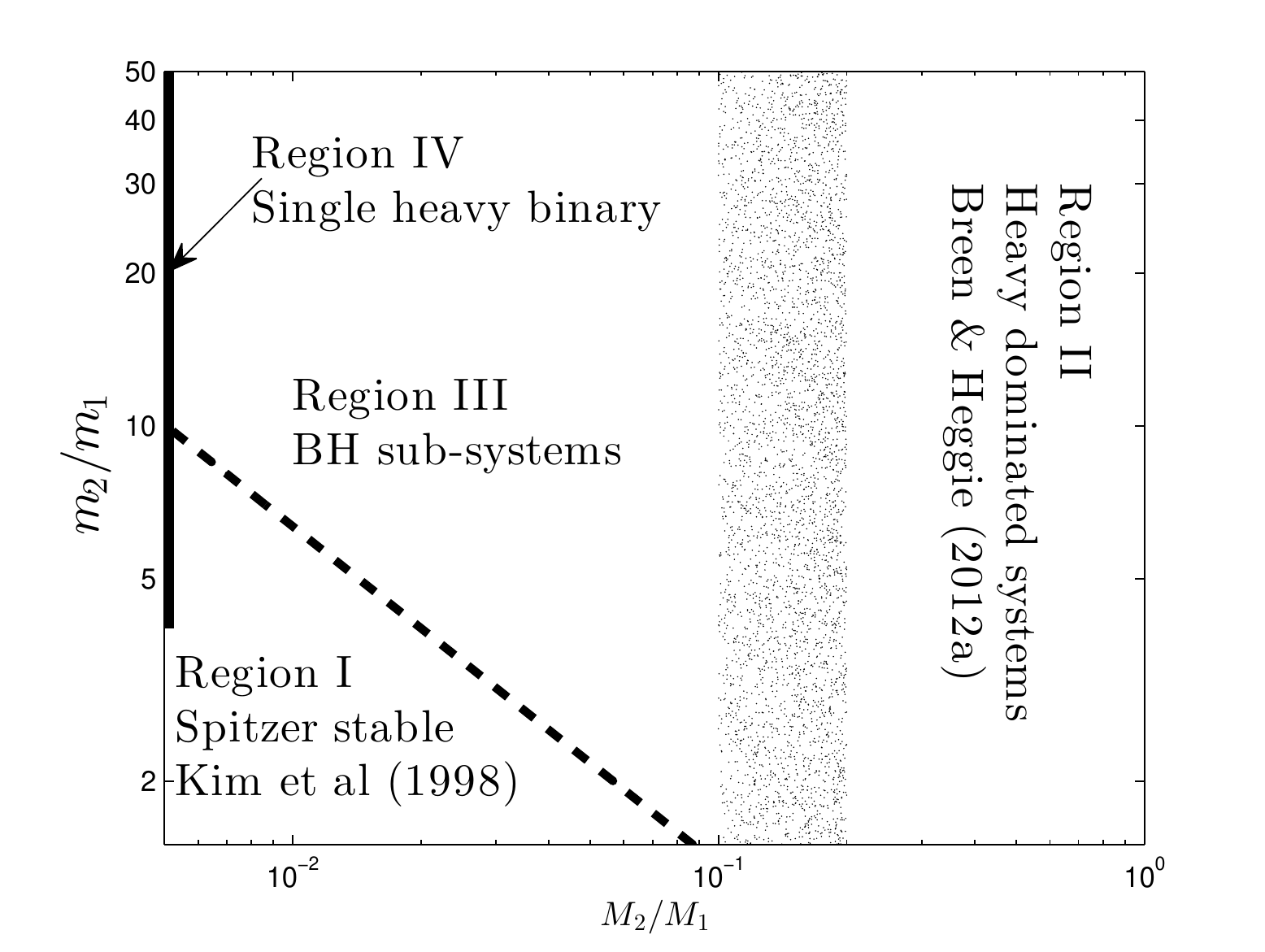}}}
\caption{The parameter space of two component systems divided up into different regions depending on their structure and the conditions for the onset   of gravothermal oscillations. See text and Table \ref{table:BHgtoregions} for details}
\label{fig:gtoparameter}
\end{figure}

\begin{table}
\caption{Summary of the different regions in the parameter space of two-component systems. The table states whether a given dynamical process is dominated by the light or heavy component of the system. $t_{\rm rh}$ represents the two-body relaxation time within $r_{\rm h}$, $t_{\rm rc}$ represents the two-body relaxation time within $r_{\rm c}$, GTO stands for gravothermal oscillations, with reference to which component contains the large isothermal region which becomes unstable, and the final column represents whether or not the system is Spitzer stable; (for Region IV, since $N_2\sim2$, Spitzer instability is not an appropriate concept). See text for further details.}
\begin{tabular}{|ccccc}
Region          & $t_{\rm rh}$  & $t_{\rm rc}$ & GTO    & Spitzer stable \\ \hline
I               & light     & heavy    & light  &  Y   \\ 
II              & heavy     & heavy    & heavy  &  N   \\ 
III             & light     & heavy    & heavy  &  N   \\ 
IV              & light     & light    & light  &  $-$   \\ 
\end{tabular}
\label{table:BHgtoregions}
\end{table}

In Section \ref{sec:GTO} we considered gravothermal oscillations in systems containing a BH sub-system. This extends the parameter space of two-component clusters studied by \cite{breenheggie1} to lower values of $M_2/M_1$ for $m_2/m_1\ge5$. But there are rather distinct physical characteristics of the systems studied in the two papers, as we have already seen in the study of gravothermal oscillations (Section \ref{sec:GTO}). Here we attempt to summarise these ideas.

In order to differentiate between two-component systems a classification scheme has been devised that divides the two-component system parameter space into four regions (see Fig. \ref{fig:gtoparameter}). The criteria used to divide the parameter space are, (a) whether or not the isothermal region which becomes gravothermally unstable is associated with the heavy component (regions II \& III) or the light component (regions I \& IV), (b) whether or not the system is Spitzer stable (region I) or Spitzer unstable (regions II \& III) and (c) whether or not the two-body relaxation process within $r_{\rm h}$ is dominated by the heavy component (region II) or the light component (regions I, III \& IV). The differences between the regions are summarised in Table \ref{table:BHgtoregions}. We shall now justify the classification by considering each of the criteria used more closely. 

The simplest distinction to make is between systems which are Spitzer stable and these which are Spitzer unstable, that is systems which achieve equipartition of kinetic energy by mass segregation and those which cannot. \cite{Spitzer} constructed the following stability condition 
$$ \frac{M_2}{M_1}  < 0.16\Big( \frac{m_2}{m_1} \Big)^{-\frac{3}{2}} $$   
based on theoretical arguments and some simplifying assumptions. However a study by \cite{Wattersetla}, using Monte Carlo simulations, found Spitzer's condition to be too strong and suggested a different condition of similar form with a different constant and power, for the range of stellar mass ratios they studied ($m_2/m_1 < 7$). For simplicity the \cite{Spitzer} condition is used in Fig. \ref{fig:gtoparameter} to divide the parameter space. The important differences between Spitzer stable and  Spitzer unstable systems are, first, that equipartition of kinetic energy holds after mass segregation (i.e. $m_2\sigma_2^2 = m_1\sigma_1^2$) and, second, that for any appreciable value of $m_2/m_1$ the value of  $M_2/M_1$ has to be significantly small. Both the systems in the present paper and the systems studied by \cite{breenheggie1} fall into the more general class of Spitzer unstable systems. Region I consists of Spitzer stable two-component systems, and occupies the lower left part of 
Fig. \ref{fig:gtoparameter}. 

Gravothermal oscillations in Spitzer stable systems were studied by \cite{KimLeeGood1998}. They argued that because of the small values of $M_2/M_1$ (and the high values of $m_2/m_1$) the heavy component was confined to the centre of the system. They showed that the systems they studied became gravothermally unstable once a certain ratio of energy flux at $r_{\rm h}$ and $r_{\rm c}$ (i.e. $\epsilon$) was reached, and that this value was the same as that for a one-component system. As the bulk of the system is in the light component this implies that the instability results from a large isothermal region in the light component. These remarks justify the entries in line 1 of Table \ref{table:BHgtoregions}.

One situation where the concept of Spitzer stability is inappropriate is where the number of heavy particles is small ($N \sim 2$). As the heavy particles tend to find their way to the core of the system and form a binary, the role of the heavy component is still significant, because this binary becomes the power source for the system lying inside the core of the light system. That is why these systems have been given their own classification, although they may be regarded as the extreme of low $M_2/M_1$ for both Spitzer stable and Spitzer unstable systems. Clearly in this case if gravothermal oscillations are found, they will result from a large isothermal region in the light component, hence the entries in line 4 of Table \ref{table:BHgtoregions}.  

Finally the last division in Fig. \ref{fig:gtoparameter} is between Regions II \& III. The space occupied by Regions II \& III consists entirely of Spitzer unstable models. The difference between these two models depends on the value of $M_2/M_1$. The case where $M_2 \ll M_1$, includes the topic of interest in the present paper (i.e. systems containing BH sub-systems). Due to small $M_2/M_1$ in these systems the light component dominates at $r_{\rm h}$ and thus the rate of energy generation is regulated by the light component. In the case where $M_2 \gtrsim 0.1M_1$ the heavy component has a significant effect on the relaxation process within $r_{\rm h}$, particularly when  $M_2 \sim M_1$ \citep{breenheggie1}.  The distinction between the two cases is clear when considering extreme values of $M_2/M_1$, but the exact division between the two is unclear and may have some dependence on $m_2/m_1$. This is why a shaded area separates the two regions in Fig. \ref{fig:gtoparameter}. In 
both cases 
the 
onset of gravothermal instability is associated with a large isothermal region in the heavy component (see Section \ref{sec:GTO} in the present paper and \cite{breenheggie1}). Another reason for the distinction between regions II \& III is the theoretical arguments given in Section \ref{sec:tidal}: if we consider tidally limited systems, $M_2/M_1$ is expected to grow with time for region II systems and decrease with time for region III systems.

\section*{Acknowledgments}
We thank S. Portegies Zwart for numerous helpful comments on a previous draft of this paper. Some of our hardware was purchased using a Small Project Grant awarded to DCH and Dr M. Ruffert (School of Mathematics) by the University of Edinburgh Development Trust, and we are most grateful for it. PGB is funded by the Science and Technology Facilities Council (STFC).

\appendix
\section{Energy transport in systems containing a BH sub-system}\label{ABH}
It has been a fundamental assumption in the present paper that, in a system containing a BH sub-system, the majority of energy generated  first flows throughout the BH sub-system and is then transferred to the rest of the system via two-body relaxation. In this section we will discuss the validity of such an assumption. We will do this by considering the energy generated by a BH binary as it hardens and is ultimately ejected from the system, as was done for the one-component case by \cite{HeggieHut2003}. We shall assume that the BH binary mostly generates energy by encounters with single BH. This is a reasonable assumption as long as there are sufficient BH for the central region to be dominated by them. We can divide the life of the BH binary in the system into five energy generating phases as follows:

\begin{enumerate}
\renewcommand{\theenumi}{(\arabic{enumi})}
\item After the BH binary is formed the interactions between the binary and the other BH will not be energetic enough to remove either from the sub-system. During this phase all the energy that is generated must be deposited within the BH sub-system.

\item After phase $(1)$ the binary then starts giving the single BH enough energy to escape the BH sub-system. The single BH will typically receive more kinetic energy from an encounter with a BH binary than the  binary itself. During this phase the BH binary remains in the BH sub-system and the increase in the kinetic energy of the centre of mass (c.m.) of the binary is deposited in the BH sub-system. Some of the energy of the single BH, however is deposited directly into the light component though how much is discussed further below.

\item At some point the c.m.  of the BH binary will receive enough kinetic energy that it too can escape from the BH sub-system along with the single BH. As the density is much lower outside the BH sub-system, however, the binary is unlikely to deposit much energy until it returns to the higher density region in the centre of the BH sub-system.

\item The single BH starts receiving enough energy to escape from the system. The BH binary still escapes from the BH sub-system but remains bound to the whole system and ultimately returns to the BH sub-system.

\item Finally the BH binary escapes from the system and the binary contributes no more energy to the system.
\end{enumerate}

Using the same approach as \citet[][ see page 225, Box 32.1]{HeggieHut2003}, we will now consider where the energy generated by each hard binary is distributed. As stated in Section \ref{sec:ej} the amount of energy generated by each hard binary is expected to be $\sim 10m_2|\phi_{\rm c}|$ and we will assume as in Section \ref{sec:ej} that $\phi_2/\phi_1 \approx 10^{-1}$. We will consider in turn heating from BH which do not escape the sub-system, from subescapers (BH which escape from the sub-system but not from the system) and finally from BH which escape the system.  

Heating from BH which do not escape the sub-system is involved in phases 1 (the single BH and the binary) and 2 (the binary). The amount of energy generated during phase 1 can be estimated in the same way as was done for the one component case by \cite{HeggieHut2003}.  The result is that only about $3\%$ of the total energy generation per hard binary is generated during this phase. Including the direct heating by the BH binary during phase 2, the total heating from BH which do not escape the sub-system is about $5.5\%$.

Heating from subescapers is involved in phases 2 (the single BH), 3 (the single BH and the binary) and 4 (the binary). The total amount of energy contributed by subescapers is about $49\%$. Initially a subescaper indirectly heats the BH sub-system, by $m_2\phi_2$ for a single BH and by $2m_2\phi_2$ in the case of the binary. The number of single subescapers is expected to be approximately $4.3$ and the number of encounters which cause the binary to become a subescaper is also approximately $4.3$. This is because in both cases $4.3$ is the typical number of encounters needed to increase the binding energy of the binary by a factor of $10$. This brings the total amount of heating (including the direct heating considered previously) to the BH sub-system to $18\%$.

What happens to a subescaper once it leaves the BH sub-system is a point of uncertainty. It will indirectly heat the light component up to some maximum radius (which depends on its kinetic energy) reached by the BH. It is possible that afterwards the subescaper remains on a nearly radial orbit and falls back into the BH sub-system, releasing most of its energy there. In the case of one-component systems \cite{Spitzer} showed that if a particle is ejected from the core its orbit is perturbed by the other stars in the system, with the result that the particle misses the core at the next pericentre of its orbit. However it is clear that the more massive BH will be less significantly perturbed by the light stars in the system, and may well return to the sub-system at the pericentre of its orbit. If the BH does return to the sub-system then most of its energy will be distributed there, and the same reasoning will also apply to non-escaping  BH binaries which are ejected from the core. If almost all the heating 
from subescapers occurs in 
the BH sub-system, then the total heating per binary (including the other heating considered previously) to the BH sub-system is now $\approx 54\%$.

Finally it is fairly easy to estimate the amount of energy resulting from escapers, which is $m_2|\phi_{\rm c}|$ per single BH escaper and $2m_2|\phi_{\rm c}|$ for the escape of the binary itself. The heating is indirect and heating to each component is proportional to the contribution of each component to the central potential. Per hard binary the percentage of energy generated due to escape is the same as for a one-component system, and is approximate $45\%$: of this about $10\%$ goes into the heating of the BH sub-system (as $\phi_{\rm c} - \phi_{\rm c,1} \simeq 0.1\phi_{\rm c}$). This type of heating is involved in phases 4 (the single BH) and 5 (the binary and single BH). Therefore the heating to the BH sub-system per hard binary may be as much as $59\%$. We have assumed (Sections \ref{sec:theory} and \ref{sec:bal}) that all energy heats the BH sub-system and is then conducted to the light component. Clearly this assumption is only approximately valid if the proportion of heating to the BH sub-system (per BH binary) is $\sim 59\%$.



\section{Heating in outer Lagrangian shells}\label{sec:Heatingouterlagra}

In this section we shall consider the heating by the BH
caused by the initial mass segregation. In simulations in
the present paper the BH are initially spread throughout
the system with the same velocity distribution as the other
stars. As the BH are much more massive than the other stars
the tendency towards equipartition of kinetic energy causes
the BH to lose kinetic energy to the other stars, which in
turn causes the BH to fall in the potential well of the cluster. 
As stated in Section \ref{sec:BHtwo}, these systems are usually Spitzer unstable,
and therefore equipartition of kinetic energy cannot be achieved: BH
continuously fall in the potential well until they are concentrated in
the centre of the system. The time this process takes depends on the
location of the BH, as the process depends on the local relaxation
time, which varies significantly throughout the cluster. For BH
which start in the outer parts of the system this process takes the
longest, as the relaxation time is longest there and they have to
travel the furthest to reach the central region.

For BH whose orbits lie mostly outside the half-mass radius of the
whole system, the loss of energy causes an increase in the mean
kinetic energy, just as for orbits in a $1/r$ potential.  Also, 
the relaxation time at the location of these BH considerably exceeds
the half-mass relaxation time.  Though the equipartition time scale is
smaller by a factor of order $m_1/m_2$, there is some radius outside
the half-mass radius at which the equipartion timescale is comparable
with the time of core collapse which, for parameters of our typical
models, can be taken to be roughly $0.3 t_\mathrm{rh}$.  Throughout
core collapse, therefore, we can expect the outermost Lagrangian radii of
the BH component to exhibit a steady rise in velocity dispersion,
while in intermediate Lagrangian shells the velocity dispersion first
increases and then decreases.  In the innermost Lagrangian shells the
velocity dispersion may be expected to decrease throughout the time to
the first core collapse.

\begin{figure}
\subfigure{\scalebox{0.60}{\includegraphics{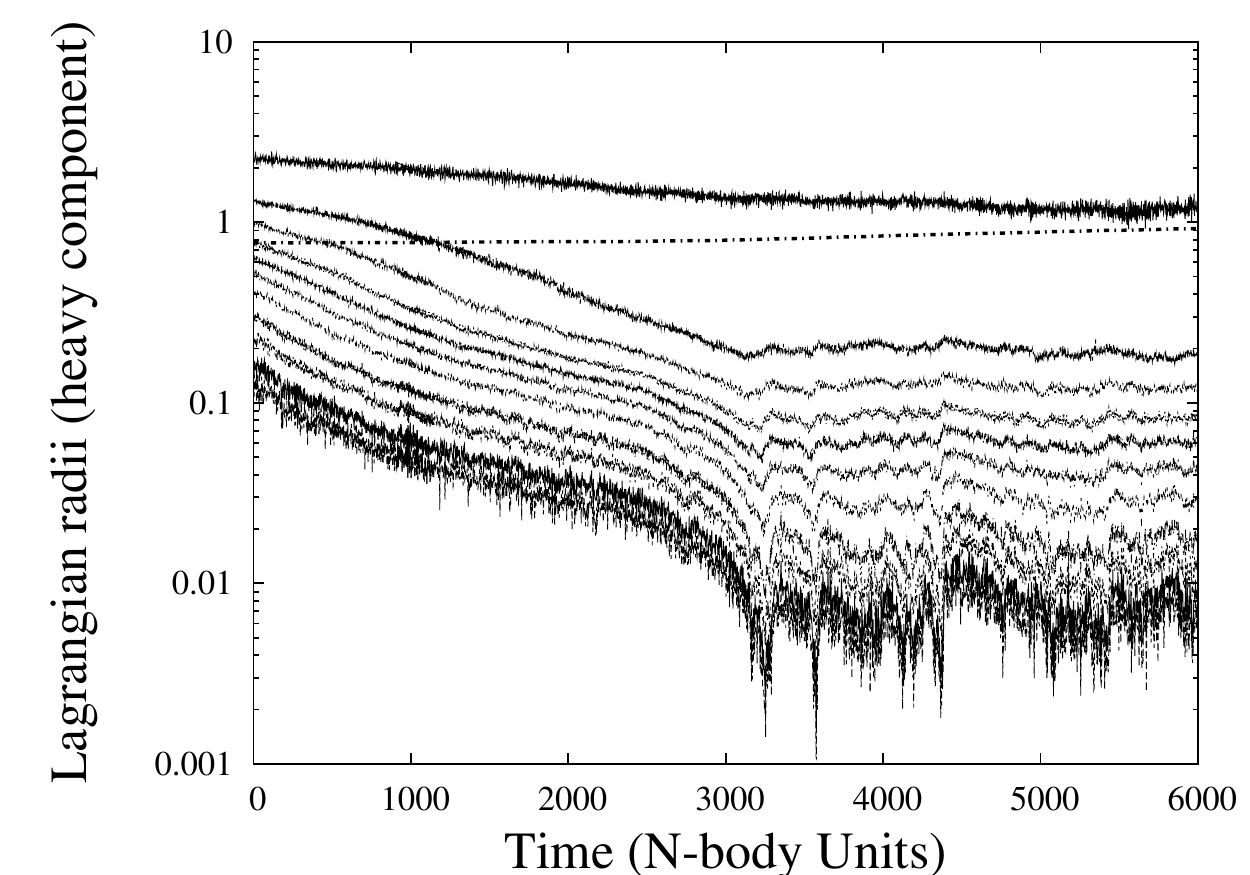}}}\quad
\subfigure{\scalebox{0.60}{\includegraphics{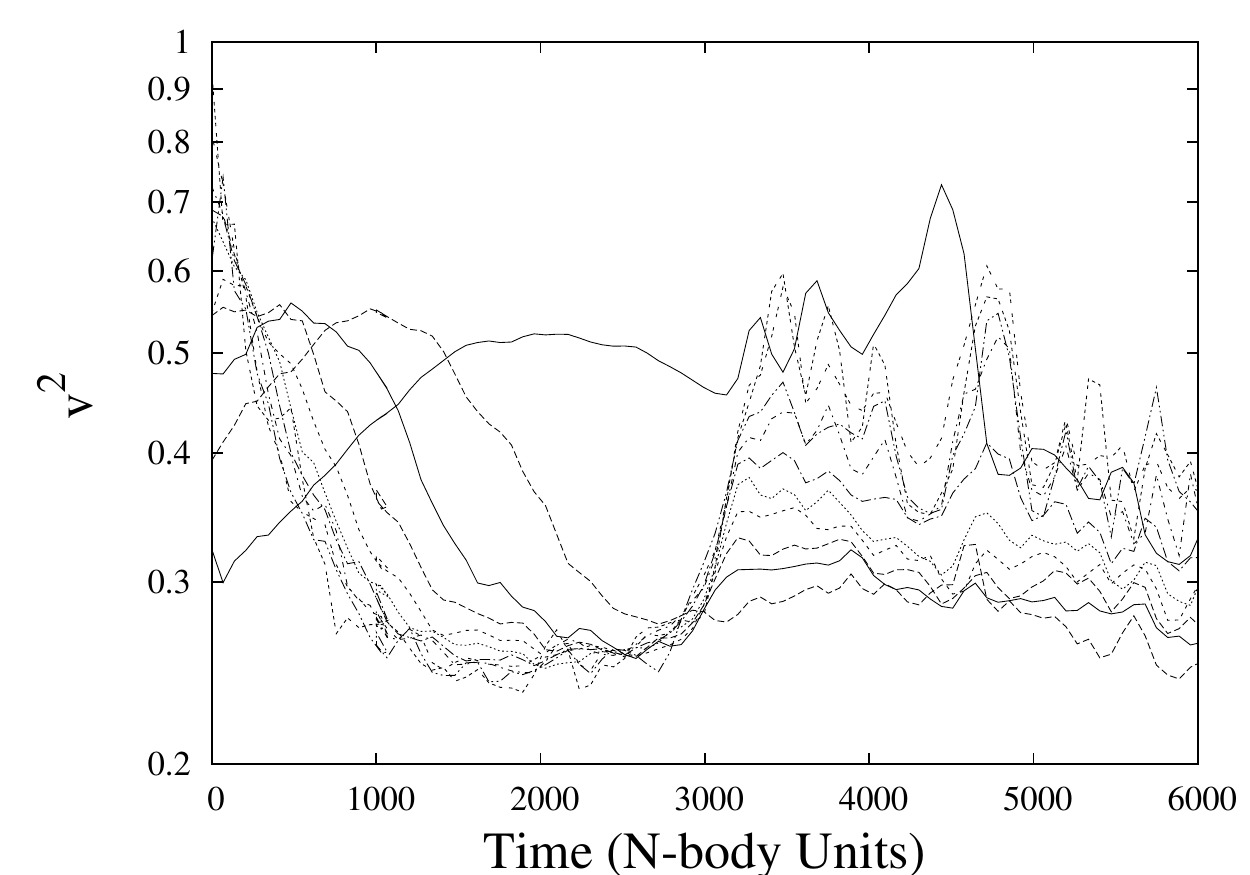}}}\quad
\caption{ Evolution of the heavy component in a two component N-body run with initial values $N=128k$, $m_2/m_1=5$ and $M_2/M_1\approx 0.038$. The corresponding mass fractions are $1\%$, $2\%$, $5\%$, $10\%$, $20\%$, $30\%$, $40\%$, $50\%$, $62.5\%$, $75\%$ and $90\%$ of $M_2$. Bottom: Mean square velocity dispersion inside the Lagrangian shell. The $90\%$ shell becomes the hottest shell at $t\simeq 1450$  and remains hotter than the next inner shell until well after $t_{\rm cc}$. Top: plot of the radii of the Lagrangian shells in the heavy component.  The dotted line is $r_{\rm h}$, the half mass radius of the whole system.}
\label{fig:NMSeg}
\end{figure}

This behaviour is illustrated in Fig.  \ref{fig:NMSeg} with an N-body
run. This run uses $m_2/m_1 = 5$, which is smaller than would be
expected for a BH sub-system; however the small value of
$m_2/m_1$
 allows for a large value of $N_2$ (in this case $N_2 = 1000$),
which more clearly illustrates the behaviour. As can be seen
in Fig. B1 (Bottom) the squared three dimensional velocity
dispersion ($v_2^2$) initially decreases for the Lagrangian radii
within $\sim r_\mathrm{h}$ and initially increases for the larger Lagrangian
radii. The heating in the outer Lagrangian radii ends 
roughly when the
radii enter $r_\mathrm{h}$ (see Fig. \ref{fig:NMSeg} Top), after which they show a decrease in $v^2$
 up until $t \approx 2500$, when there is an increase in $v^2$
which is associated with core collapse. By the time that core
collapse has completed (at $t\simeq 3250$) most of the BH (80\%)
have been segregated to within $\approx 0.2r_\mathrm{h}$.  At this time the
outermost Lagrangian radius (90\% of $M_2$) has the highest
value of $v^2$
(Fig.  \ref{fig:NMSeg} Bottom), though part
of the increase may be attributable to BH ejection.  The outermost BH are still undergoing mass segregation.  Indeed
as can be seen in Fig. \ref{fig:NMSeg} (Top) the outermost Lagrangian
radius continuously contracts even after core collapse occurs.

The behaviour discussed in this section is not limited to systems containing 
a BH  sub-system; in fact any system with a mass spectrum should also exhibit similar 
behaviour. The important point is that continued mass segregation serves to 
heat the low-mass component, and this kind of heating may lower the required rate of energy generation from the core, as energy is directly injected 
into the region near $r_{\rm h}$. This effect has not been taken into account 
in the theory in Section \ref{sec:BHtwo}, but as the values of $N_2$ are so 
small in Section \ref{sec:BHtwo} it is unlikely that this effect is significant.

\bsp

\label{lastpage}

\end{document}